\documentclass[aps,prd,showpacs,amssymb,floatfix,nofootinbib]{revtex4-1}
\usepackage{graphicx,amsmath,subfigure,array}

\newcommand{\eeq}{\end{equation}}
\newcommand{\bea}{\begin{eqnarray}}

\def\ltsima{$\; \buildrel < \over \sim \;$}
\def\simlt{\lower.5ex\hbox{\ltsima}}
\def\gtsima{$\; \buildrel > \over \sim \;$}
\def\simgt{\lower.5ex\hbox{\gtsima}}
\newcommand{\hGR}{\mathbf{h}_\mathrm{GR}}
\newcommand{\hAP}{\mathbf{h}_\mathrm{AP}}
\newcommand{\tbf}{\theta}
\newcommand{\ttr}{\hat \theta}

\def\ltsima{$\; \buildrel < \over \sim \;$}
\def\simlt{\lower.5ex\hbox{\ltsima}}
\def\gtsima{$\; \buildrel > \over \sim \;$}
\def\simgt{\lower.5ex\hbox{\gtsima}}

\def\lesssim{\mathrel{\hbox{\rlap{\hbox{\lower4pt\hbox{$\sim$}}}\hbox{$<$}}}}
\def\gtrsim{\mathrel{\hbox{\rlap{\hbox{\lower4pt\hbox{$\sim$}}}\hbox{$>$}}}}
\def\alt{\mathrel{\hbox{\rlap{\hbox{\lower4pt\hbox{$\sim$}}}\hbox{$<$}}}}
\def\agt{\mathrel{\hbox{\rlap{\hbox{\lower4pt\hbox{$\sim$}}}\hbox{$>$}}}}
\def\PRD{{\it Phys. Rev.} D~}
\def\PRL{{\it Phys.Rev.} Lett~}

\def\CQG{{\it Class. Quantum Grav.}}

\def\gta{\ifmmode {\mathbin{\lower 3pt\hbox   
    {$\,\rlap{\raise 5pt\hbox{$\char'076$}}\mathchar"7218\,$}}}
    \else {${\mathbin{\lower 3pt\hbox
    {$\rlap{\raise 5pt\hbox{$\char'076$}}\mathchar"7218\,$}}}
    $}\fi}
\def\lta{\ifmmode {\,\mathbin{\lower 3pt\hbox   
    {$\,\rlap{\raise 5pt\hbox{$\char'074$}}\mathchar"7218\,$}}}
    \else {${\mathbin{\lower 3pt\hbox
    {$\rlap{\raise 5pt\hbox{$\char'074$}}\mathchar"7218\,$}}}
    $}\fi}

\begin{document}
\title{Influence of conservative corrections on parameter estimation for EMRIs}
\author{E. A. Huerta and Jonathan R. Gair}

\affiliation{Institute of Astronomy, Madingley Road, CB3 0HA Cambridge, UK}

\email{eah41@ast.cam.ac.uk}
\email{jgair@ast.cam.ac.uk}


\date{\today}

\begin{abstract}
We present an improved numerical kludge waveform model for circular, equatorial extreme-mass-ratio inspirals (EMRIs). The model is based on true Kerr geodesics, augmented by radiative self--force corrections derived from perturbative calculations, and in this paper for the first time we include conservative self-force corrections that we derive by comparison to post-Newtonian results. We present results of a Monte Carlo simulation of parameter estimation errors computed using the Fisher Matrix  and also assess the theoretical errors that would arise form omitting the conservative correction terms we include here. We present results for three different types of system, namely the inspirals of black holes, neutron stars or white dwarfs into a supermassive black hole (SMBH). The analysis shows that for a typical source (a \(10 M_{\odot}\) compact object captured by a \(10^6 M_{\odot} \) SMBH at a signal to noise ratio of 30) we expect to determine the two masses to within a fractional error of \(\sim 10^{-4}\), measure the spin parameter \(q\) to \(\sim 10^{-4.5}\) and determine the location of the source on the sky and the spin orientation to within \(10^{-3}\) steradians. We show that, for this kludge model, omitting the conservative corrections leads to a small error over much of the parameter space, i.e., the ratio \(\cal{R}\) of the theoretical model error to the Fisher Matrix error is \( \cal{R}\) \( <1 \) for all ten parameters in the model. For the few systems with larger errors typically \(\cal{R}\) \(< 3\) and hence the conservative corrections can be marginally ignored. In addition, we use our model and first order self--force results for Schwarzschild black holes to estimate the error that arises from omitting the second-order radiative piece of the self-force. This indicates that it may not be necessary to go beyond first order to recover accurate parameter estimates.  
\end{abstract}

\pacs{}

\maketitle

\section{Introduction}    
One of the most important sources of gravitational waves (GWs) for the  Laser Interferometer Space Antenna (LISA) will be  the extreme-mass-ratio inspirals (EMRIs) of stellar mass compact objects into massive black holes (BHs). These systems occur in the centres of galaxies where massive black holes are surrounded by clusters of stars which contain large numbers of compact stellar remnants. The heavier black holes sink toward the centre as a result of mass segregation during random encounters among stars. These encounters will gradually perturb the orbits until one of the stellar mass black holes passes close to the central black hole. If the impact parameter is sufficiently small, enough orbital energy will be radiated in gravitational waves to leave the remnant bound to the BH. Thereafter, the small black hole orbit gradually decays via GW emission.  Eventually the black hole may end up on an orbit where it is radiating GWs continuously in the LISA band. Even then, the black hole may complete several hundred thousand orbits before crossing the horizon.

The GWs emitted by these sources will be rich with information. They will provide a map of the space-time exterior to the large body and the  response of the horizon to tidal forces. Furthermore, tests of the no-hair theorem will be possible as LISA will be able to measure the black hole's mass, spin and quadrupole moment to  fractional accuracies of \(10^{-3}\) \cite{ryan}. Testing this theorem will allow us for the first time to confirm or falsify the predictions that general relativity makes about black hole solutions. On the other hand, we could discover non--black hole systems in nature, e.g., boson stars or naked singularities. Additionally, the accurate system masses and spins that LISA should determine will be of great importance for understanding the stellar populations in the central parsecs of galactic nuclei.  

In contrast to comparable mass binary systems, where modelling may be done using post--Newtonian (PN) theory in the early inspiral and numerical relativity for the final orbits and merger, EMRI GWs may be obtained  using Black Hole Perturbation Theory (BHPT). This is because  the extreme mass ratio, \(\eta= m/M\), can be used as a small expansion parameter.  Such EMRIs will be visible to LISA for central black holes with masses in the range  \(10^{4}-10^7\) \(M_{\bigodot}\) and out to redshifts \(z \approx 1\) \cite{seoane,lrs}. As we will show later, the small mass ratio \(\eta  \) has an important effect on the number of orbits that the small body performs before merging.

Detection of EMRIs will be difficult, since it is expected that the signal from EMRIs will be about an order of magnitude below LISA's projected instrumental noise. In fact, at lower frequencies, below 2 mHz, the signal will be orders of magnitude below the confusion noise due to unresolved galactic binaries. However, signals can be identified in noisy data using matched filtering. This requires the use of theoretical templates whose  phase remains accurate to one cycle over the \(\eta^{-1}\) cycles of the waveform generated while the orbit of the stellar mass compact object is in the strong curvature region, close to the large black hole. Hence, there is a pressing need for accurate theoretical models of the waveforms. Modelling this emission is a difficult problem, although there has been great progress in recent years. 

BHPT relies upon the fact that the stellar mass compact object can be treated as a small perturbation of the gravitational field of the central black hole. Over short timescales, the compact object follows a Kerr space-time geodesic since we can neglect back--reaction for that short stretch of time. A Kerr geodesic is characterized by  three constants of motion, namely the energy, \(E\), the angular momentum about the hole's spin axis \(L_{z}\), and the Carter constant \(Q\), which is a relativistic generalization of the third integral of motion, which arises from the separation of the equations of motion for orbits in an axisymmetric gravitational potential. In the spherical limit, i.e., for a non-rotating central black hole, \(Q\) reduces to the square of the angular momentum projected into the equatorial plane. Over longer timescales, \(\sim M/ \eta\), radiation--reaction causes the orbit to evolve adiabatically. This can be characterized by changes in the geodesic orbital elements \(E,L_{z}\) and \(Q\). As the compact object inspirals, it eventually passes through an innermost stable orbit where adiabaticity breaks down. It then  follows a geodesic plunge orbit and is swallowed by the central black hole. 

First order radiative BHPT in a Kerr background is described by the Teukolsky formalism. Unfortunately, Teukolsky--based waveforms are very computationally expensive to generate. However, there exist various families of approximate waveforms that capture the main features of true signals and are much quicker to generate. One family are the post--Newtonian (PN) waveforms, which are both analytic and easy to generate. Recently, a class of PN waveforms were developed for the EMRI case by Barack and Cutler \cite{cutler}. These GWs were constructed considering the lowest--order quadrupole waveforms for eccentric binaries on Keplerian orbits derived by Peters and Matthews \cite{peters}, but the orbits were corrected to include the effects of pericenter precession, Lense--Thirring precession, and inspiral due to radiation reaction. This family of GWs are useful for scoping out data analysis issues, e.g., computing the Fisher information matrix to estimate  parameter measurement accuracies. However, a small black hole in a close orbit about a much larger one is too relativistic a system for PN analyses to be valid. The PN expansion is unlikely to be reliable in the EMRI limit since the object spends so much time in the strong field region of the space-time where \(v \sim c\). But, it is in this zone where most of the GWs observable to LISA are generated. In order to bridge this gap, we again rely upon BHPT. 

Another family of EMRIs waveforms are referred to as ``numerical kludge'' (NK)\cite{kludge}. These are constructed by combining an exact particle trajectory with a flat space--time wave--emission formula. This framework also aims to capture the main features of the waveform accurately. Comparison to Teukolsky--based waveforms indicates that NK waveforms for geodesic orbits are quite faithful as the overlap between the two types of waveforms for geodesic orbits is greater than \(0.95\) over a considerable portion of the inspiral parameter space \cite{kludge}. However these models are incomplete in their treatment of the orbital evolution under the self--force since the model currently does not include conservative radiation--reaction terms, and the waveform--emission formula and phase space trajectories are approximate. 

There is currently some debate in the literature \cite{pound,flan} as to the importance of the conservative pieces of the self--force for waveform modelling, both in terms of source detection, and parameter estimation. The purpose of this paper is to shed some light on this issue using the numerical kludge waveform family. We will describe how to include conservative corrections in our model, and then estimate their influence on parameter extraction using the formalism recently developed by Cutler and Vallisneri \cite{vallisneri}.

This paper is organized as follows. In section~\ref{s2} we construct asymptotic observables for our kludge waveform model, namely the orbital frequency and its first time derivative. We show how we can effectively  include both components of the self--force: radiative and conservative, to second and first order, respectively, by comparison to PN waveforms. We also discuss the relative importance of the various conservative corrections to the phase evolution of our waveform signal.  In section~\ref{s3} we use our waveform model to assess the importance of the second order radiative piece that will be missing once first order accurate self--force waveforms are generated. We do this by comparing to  self--force data obtained by Barack and Sago \cite{sago} for circular equatorial orbits around a Schwarzschild black hole. In  section~\ref{s4} we will use our inspiral kludge waveforms including modulations from LISA's response function to estimate the  noise--induced parameter errors using the Fisher Matrix formalism. These provide confirmation and extension of the results obtained using the analytic kludge model of Barack and Cutler~\cite{cutler}. In  section~\ref{s5} we employ the recently developed formalism of Cutler and  Vallisneri \cite{vallisneri}  to estimate the theoretical parameter errors that would be introduced by  waveform template inaccuracies. Specifically, we study the importance of the conservative corrections by computing the theoretical errors in the parameters that would arise  from omitting conservative corrections from the model. Finally, we summarize our work in section~\ref{s6}.

\section{Kludge waveform}
\label{s2}
Our waveform is based on the numerical kludge approach described elsewhere~\cite{improved,kludge}, but we summarise the formalism here. The aim is to capture the main features of the waveform inspiral in the strong field regime. Because an accurate description of the orbital and phase evolution of the small body is not yet available, we try to circumvent this problem in various steps. The system's extreme mass ratio \(m/M \sim 10^{-6}\) guarantees that gravitational back--reaction effects occur on timescales much longer than any orbital timescale. Hence, we can assume that the inspiralling object instantaneously follows a Kerr space-time geodesic. We slowly evolve the parameters of the geodesic using a prescription for the rate of change of energy, angular momentum and Carter constant and calculate the resulting trajectory of the inspiralling body in the Boyer--Lindquist coordinates of the Kerr space-time of the central black hole. The gravitational waveform is then constructed by identifying these coordinates with spherical polar coordinates in a pseudo-flat space and using weak-field emission formulae.

In this paper, we restrict our attention to circular equatorial orbits. For such orbits, the inclination angle and eccentricity remain constant, i.e., circular-equatorial orbits remain circular-equatorial~\cite{ori}. The energy  \(E\) and angular momentum \(L_{z}\) of such orbits is given in terms of the Boyer-Lindquist radius of the orbit, \(p\), as~\cite{chandra} 
\begin{eqnarray}
\frac{E}{m} &=& \frac{1 - 2\left(M/p\right) \pm \left(a/M\right)\left(M/p\right)^{3/2}}{\sqrt{1 - 3\left(M/p\right) \pm 2\left(a/M\right)\left(M/p\right)^{3/2}}}, \nonumber \\   \frac{L_{z}}{m M} &=& \pm \left(\frac{p}{M}\right)^{1/2}\frac{1  \mp 2\left(a/M\right)\left(M/p\right)^{3/2} + \left(a/M\right)^{2}\left(M/p\right)^{2 }}{\sqrt{1 - 3\left(M/p\right) \pm 2\left(a/M\right)\left(M/p\right)^{3/2}}}, \label{1.1}
\end{eqnarray}
\noindent  where the upper (lower) sign is for prograde (retrograde) orbits and \(a\) stands for the spin of the central black hole. From now on we will write \(q=a/M\). The Carter constant, $Q=0$, for equatorial orbits. 

To obtain the orbital evolution of the compact object, we need to calculate the evolution of the energy  \(E\) and angular momentum \(L_{z}\). We can evaluate these quantities by equating their rate of change with the flux carried away by the gravitational waves, \(\dot{E}\) and \(\dot{L_{z}}\). We will use the radiation fluxes derived by Gair \& Glampedakis~\cite{improved} to compute the compact object's orbital evolution. These fluxes are based on 2.5PN expressions derived by Tagoshi \cite{tagoshi}, augmented by fits to accurate BHPT results and with various consistency conditions imposed to ensure physical behaviour for near-circular and near-polar orbits. 

To evolve a circular orbit  we need only to specify the angular momentum or the energy flux, as they are related by the `circular goes to circular' rule~\cite{ori}
\begin{equation}
\dot E(p) = \pm \frac{\sqrt{M}}{p^{3/2} \pm a \sqrt{M}} \dot L_{z}(p) = \Omega(p) \dot L_{z}(p),
\label{1.3}
\end{equation}
\noindent where \( \mathrm{d} \phi / \mathrm{d} t = \Omega(p)\), is the azimuthal velocity of the orbit. We choose to work with the evolution of \(L_{z}\), which has the following form~\cite{improved}
\begin{eqnarray}
\dot{L}_z&=& -\frac{32}{5} \frac{m^2}{M} \left(\frac{M}{p}\right)^{7/2} 
\Bigg\{ 1 -\frac{61}{12}q\left(\frac{M}{p}\right)^{3/2} - \frac{1247}{336}\left(\frac{M}{p}\right)  +
4\pi \left(\frac{M}{p}\right)^{3/2}   \nonumber \\ &&  - \frac{44711}{9072}\left(\frac{M}{p}\right)^2  
  + \frac{33}{16}\,q^2\left(\frac{M}{p}\right)^2  + \textrm{higher order Teukolsky fits}\Bigg\} .\
\label{new_Ldot}
\end{eqnarray}

\noindent The `higher order Teukolsky fits' are given in~\cite{improved}. We do not give these explicitly here, as they are not needed to derive the conservative corrections. However, we will include them to evolve orbits when we generate waveforms.

The evolution in time of the radial coordinate is given by
\begin{equation}
\label{6}
\dot p= \frac{\mathrm{d} p}{\mathrm{d} E}\dot E= \frac{\mathrm{d} p}{\mathrm{d} L_z}\dot L_z .
\end{equation}
  
\noindent Using the exact geodesic expression for  $\mathrm{d} p/\mathrm{d} L_z$ generates inspirals that are closer to Teukolsky based evolutions than expanding the above expression at 2PN order. However, we will need the 2PN expression in the following, which is
 \begin{eqnarray}
\label{7}
\frac{\mathrm{d}p}{\mathrm{d}t} &=& - \frac{64}{5}\eta \left(\frac{M}{p}\right)^{3}\Bigg\{1- \frac{743}{336}\left(\frac{M}{p}\right) + \left(4 \pi - \frac{133}{12} q \right)\left(\frac{M}{p}\right)^{3/2} \nonumber\\ &+& \left(\frac{34103}{18144} + \frac{81}{16} q^{2}\right)\left(\frac{M}{p}\right)^{2}\Bigg\}.
\end{eqnarray}

\noindent As before \(\eta = m/M\). Up to this point our analysis has been incomplete as we have only considered the three integrals of motion. There are also three positional constants of the motion, which basically label the position of the test particle along the geodesic trajectory at some fiducial time. The evolution of these is non--trivial because  the self--interaction between the inspiralling body and the central black has two main pieces. The only one we have considered so far is the dissipative or radiative self--force and affects the evolution of the orbit. However, in real inspirals there is a second piece called the conservative self--force, which affects the principal constants of the motion. In other words, it changes the frequency of an orbit at a given radius but does not cause the orbit to evolve. We need to include this piece of the self--force as it could lead to several cycles of phase discrepancy in our kludge waveform over the inspiral.

The conservative self--force has two parts. One part is oscillatory and averages to zero, whereas the other piece accumulates and affects  the phasing of the waveform over time. In our kludge we may account for this effect by  changing  the \(\phi\) frequency. Specifically, we include this effect by writing
\begin{equation}
\frac{\mathrm{d}\phi}{\mathrm{d}t} = \left(\frac{\mathrm{d}\phi}{\mathrm{d}t}\right)_{\mathrm{geo}}\bigg(1+  \delta \Omega \bigg).
\label{7.1}
\end{equation}

\noindent This equation includes the phase derivative for a geodesic, labeled by the subscript ``geo'', and a frequency shift which will depend on the instantaneous orbital parameters. A problem arises here because to compute the necessary frequency shifts within our framework, i.e., BHPT, would require self--force calculations. At present, these are known only for particles moving on circular geodesic orbits around a Schwarzschild black hole, and only in a particular gauge \cite{sago}. But, there is an ongoing effort to extend such calculations to generic inspiral orbits in a Kerr background. The main challenge in this effort has to do with  gauge freedoms as we will discuss later.

However, we do know conservative corrections in the post--Newtonian framework up to 2PN order which include spin--orbit, spin--spin effects and finite mass contributions  \cite{blanchet}. We can combine these expressions with the radiative self--force obtained  by Tanaka et al. \cite{tanaka} based on the Teukolsky and Sasaki--Nakamura formalisms for perturbations around a Kerr BH, which includes terms of order \(q^{2}\). Equipped with these results  we will extend the method proposed by Babak et al. \cite{kludge} to include conservative corrections in the kludge model. They computed the 1PN conservative correction for circular orbits in the Schwarzschild space-time, and we now extend that calculation to derive 2PN conservative corrections for circular orbits in the Kerr space-time.

The idea is to correct the kludge expressed in a particular coordinate system, in order to ensure that asymptotic observables are consistent with post--Newtonian results in the weak field. In particular, we aim to modify the orbital frequency and its first time derivative. By modifying these two quantities, we can both  identify coordinates between the two formalisms and find the missing conservative pieces.  

From equation \eqref{1.3}, at 2PN order the orbital frequency takes the form 
\begin{equation}
\label{eq.5}
\Omega = \frac{1}{M}\left(\frac{M}{p}\right)^{3/2}\left(1- q \left(\frac{M}{p}\right)^{3/2} + O\left(\frac{M}{p}\right)^{5/2}\right),
\end{equation}
\noindent which we now augment by including conservative corrections 
\begin{eqnarray}
\frac{\mathrm{d}\phi}{\mathrm{d}t} \equiv \Omega &=&\frac{1}{M}\left(\frac{M}{p}\right)^{3/2}\left(1- q \left(\frac{M}{p}\right)^{3/2}\right) \bigg(1 + \delta \Omega \bigg),  \nonumber\\ 
 & =& \frac{1}{M}\left(\frac{M}{p}\right)^{3/2}\left(1- q \left(\frac{M}{p}\right)^{3/2}\right)\Bigg\{1 +  \nonumber\\ &+&\eta \left( d_0 + d_1 \left(\frac{M}{p}\right)+ (d_{1.5} + q\, f_{1.5})\left(\frac{M}{p}\right)^{3/2} + d_2\left(\frac{M}{p}\right)^{2}\right)\Bigg\}.\label{omCC}
\end{eqnarray}
 \noindent We will use this expansion for \(\Omega_{\rm geo}\) only to derive the conservative corrections. As with $\mathrm{d} p/\mathrm{d} L_z$, it has been shown that more reliable waveforms can be obtained by including the full geodesic frequency where it is known, c.f., \eqref{7.1}, and this will be the approach used in section~\ref{s4}. It is inconsistent to include some pieces of the evolution at arbitrary PN order, while including the conservative corrections only at 2PN. However, the lower order effects that we are including with greater accuracy do have a more significant impact on the waveform. Moreover, it has been found in the past that including higher order terms where they are known is the right strategy to obtain accurate waveforms~\cite{improved}.
 
It is worth noting that the expansion in $p$ is an expansion in \(v^{2}= M/p\). To derive the conservative corrections, we choose to leave the time derivative of the radial coordinate unchanged and given by equation \eqref{6}/\eqref{7}. This amounts to a choice of gauge such that the $\eta^2$ piece of ${\rm d}p/{\rm d}t$ vanishes. Differentiation of~(\ref{omCC}) then gives \(\mathrm{d} \Omega/ \mathrm{d} t\) for the kludge, 
\begin{eqnarray}
\frac{\mathrm{d} \Omega}{ \mathrm{d} t} &=& \frac{96}{5}\frac{\eta}{M^2}\left(\frac{M}{p}\right)^{11/2}\Bigg\{ 1+ \eta\, d_0 + \frac{M}{p} \left(-\frac{743}{336} + \eta \left(\frac{5}{3} d_1 - \frac{743}{336} d_0 \right)\right) + \nonumber\\  &+& \left(\frac{M}{p}\right)^{3/2}\left(4 \pi - \frac{157}{12} q + \eta \left(4 \pi d_0 + 2 d_{1.5} + q \left(2 f_{1.5}- \frac{157}{12}d_0\right)\right)\right) \nonumber \\ &+& \left(\frac{M}{p}\right)^{2}\left(\frac{34103}{18144}+ \frac{81}{16}q^2 + \eta\left(\frac{34103}{18144}d_0 + \frac{81}{16}q^2 d_0 - \frac{3715}{1008}d_1+ \frac{7}{3} d_2\right)\right) \Bigg\}.\nonumber\\
\label{odot}
\end{eqnarray}
We may now write down a coordinate transformation  to relate our coordinates with those used in the post--Newtonian formalism.
\begin{eqnarray}
p &=& R \Bigg\{ 1 + \left(\frac{M}{R}\right)b_1 + \left(\frac{M}{R}\right)^{3/2}(b_{1.5} + q \,s_{1.5}) + \left(\frac{M}{R}\right)^{2}b_2 \nonumber\\  &+& \eta \left( c_0 +\left(\frac{M}{R}\right)c_1 + \left(\frac{M}{R}\right)^{3/2}(c_{1.5} + q\, g_{1.5}) + \left(\frac{M}{R}\right)^{2} c_2\right)\Bigg\}, \label{coord}
\end{eqnarray}
\noindent where \(R\) denotes the post--Newtonian semi-major axis. We can now substitute this expression for the coordinate transformation into relations \eqref{omCC} and \eqref{odot}. 

The final stage of the computation is to compare the expressions for \(\Omega\) and \(\dot \Omega\), where a dot denotes \(\mathrm{d}/ \mathrm{d} t\),  with the available post--Newtonian expansions. The post--Newtonian expansions are available to higher order in the mass ratio \(\eta\), but we keep \(\eta\) only to the same order  as the kludge, Eq. \eqref{odot}. The post--Newtonian expressions for the orbital frequency and its first time derivative are given by 
\begin{eqnarray}
\Omega_{PN}^{2} &=& \frac{m_{\textrm{T}}}{R^{3}}\Bigg\{1- \frac{m_{\textrm{T}}}{R}\left(3 - \eta \right) - \left( \frac{m_{\textrm{T}}}{R}\right)^{3/2}\left(2 \left(\frac{M}{m_{\textrm{T}}}\right)^{2} + 3 \eta \right) \boldsymbol{\hat L} \cdot \mathbf{q}\nonumber\\ &+& \left( \frac{m_{\textrm{T}}}{R}\right)^{2} \left( 6 + \frac{41}{4} \eta\right)\Bigg\},
\label{1}  
 \end{eqnarray}
\noindent where \(m_{\textrm{T}} = M + m \) and \( \boldsymbol{\hat L}\) is a unit vector directed along the orbital momentum. Additionally,        
\begin{eqnarray}
\dot \Omega_{PN} &=& \frac{96}{5}\eta m_{\textrm{T}}^{5/3}\omega^{11/3} \Bigg\{ 1- \left( \frac{743}{336} + \frac{11}{4} \eta \right)(m_{\textrm{T}} \omega)^{2/3} + (4 \pi - \beta)(m_{\textrm{T}}\omega) \nonumber \\ &+& \left( \frac{34103}{18144} + \frac{81}{16} q^{2} + \eta \left( \frac{13661}{2016} + \zeta q^{2}\right) \right) (m_{\textrm{T}} \omega)^{4/3} \Bigg\},
\label{2}
\end{eqnarray}
\noindent where the spin--orbit parameter \(\beta\) is given by \(\beta = 1/12 \left( 113 M^2/ m_{\textrm{T}}^2 + 75 \eta \right)  \boldsymbol{\hat L} \cdot \mathbf{q}\) and the constant \(\zeta\) will be determined from the kludge prescription. Note that we have assumed that the spin of the inspiralling black hole is negligible with respect to the central one and so we ignore spin--spin interactions \cite{seoane}. 

We now rewrite these expressions in a convenient way to take the EMRI limit, by writing \(m_{\textrm{T}} = M (1+ \eta)\). We find,
 \begin{eqnarray}
\label{3} 
\Omega_{PN} &=& \frac{1}{M}\left(\frac{M}{R}\right)^{3/2}\Bigg\{1 + \frac{\eta}{2}- \frac{M}{R}\left(\frac{3}{2} + \frac{7}{4} \eta \right) - \,q  \left( \frac{M}{R}\right)^{3/2}\left(1+ \frac{3}{2} \eta \right) \nonumber \\ &+& \left( \frac{M}{R}\right)^{2} \left( \frac{15}{8} + \frac{169}{16} \eta\right)\Bigg\},
\end{eqnarray}

\begin{eqnarray}
\label{4}
\dot \Omega_{PN} &=& \frac{96}{5}\frac{\eta}{M^2} \left(\frac{M}{R}\right)^{11/2} \Bigg\{ 1 + \frac{3}{2}\eta- \left( \frac{2591}{336} + \frac{13571}{672} \eta \right)\frac{M}{R} \nonumber \\ &+& \left(4 \pi - \frac{157}{12}q + \eta \left(12 \pi - \frac{149}{6}q\right)\right)\left(\frac{M}{R}\right)^{3/2} \nonumber \\ &+& \left( \frac{22115}{648} + \frac{81}{16} q^{2} + \eta \left( \frac{87044}{567} +  q^{2}\left(\frac{567}{32}+ \zeta\right) \right)\right) \left(\frac{M}{R}\right)^{2} \Bigg\}.\nonumber\\ \end{eqnarray}

\noindent A direct comparison between the two expressions for the orbital frequencies and their first time derivatives allow us to solve simultaneously for \(b_1,\, b_{1.5},\, s_{1.5}, \,c_0, \,c_1\), \(c_{1.5}, \,g_{1.5}\),\,\(c_2, \,d_0\), \(\,d_1\), \(d_{1.5}, \,f_{1.5},\) and \(d_2\). We find that the non--vanishing parameters are

\begin{eqnarray}
b_1 = 1, \qquad c_0 =- \frac{1}{4}, \qquad c_1 = \frac{845}{448}, \qquad d_0 = \frac{1}{8}, \qquad d_1 = \frac{1975}{896} \nonumber\\
c_{1.5}= -\frac{9}{5}\pi, \qquad d_{1.5}= -\frac{27}{10} \pi, \qquad f_{1.5}= -\frac{191}{160}, \qquad g_{1.5}= -\frac{91}{240}\nonumber\\
 c_2 = -\frac{2 065 193}{677 376}, \qquad d_2 = \frac{1 152 343}{451 584}. \qquad \qquad \qquad
\label{10}
\end{eqnarray}

\noindent These parameters not only give us the missing conservative pieces, but also the value of the constant \(\zeta\) of equation \eqref{4} and the coordinate transformation. We find that \(\zeta = - 243/32\) and hence, the first time derivative of the post--Newtonian orbital frequency at 2nd order including spin effects and conservative corrections is given by

\begin{eqnarray}
\label{eq.11}
\dot \Omega_{PN} &=& \frac{96}{5}\frac{\eta}{M^2} \left(\frac{M}{R}\right)^{11/2} \Bigg\{ 1 + \frac{3}{2}\eta- \left( \frac{2591}{336} + \frac{13571}{672} \eta \right)\frac{M}{R} \nonumber\\ &+& \left(4 \pi - \frac{157}{12}q + \eta \left(12 \pi - \frac{149}{6}q\right)\right)\left(\frac{M}{R}\right)^{3/2} \nonumber \\ &+& \left( \frac{22115}{648} + \frac{81}{16} q^{2} + \eta \left( \frac{87044}{567} + \frac{81}{8} q^{2} \right)\right) \left(\frac{M}{R}\right)^{2} \Bigg\}.
\end{eqnarray}

\noindent Having obtained the conservative pieces, we can now assess how important they are in determining the phasing of the waveform. First, we will look at the number of cycles that the orbiting stellar--mass compact object performs in the last year of inspiral before plunge, and how many of these are contributed by each term in \(\Omega\). We focus on this particular period of time as that is when most of the GWs are radiated in the LISA band.

\begin{figure}
\centering
\includegraphics{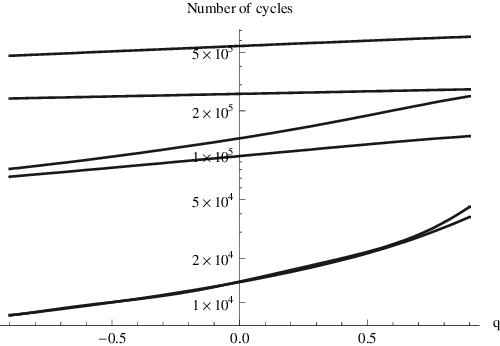}
\caption{This figure shows the number of gravitational waveform cycles generated by the inspiral of a stellar--mass compact object of specific mass \(\bar{m}\) falling into a central Kerr black hole of specific mass \(\bar{M}\) over the last year before plunge. The number of cycles is derived including conservative corrections up to order 2PN. We show results for both prograde ($q>0$) and retrograde ($q<0$) orbits for the following binary systems, \(\{\bar{M},\bar{m}\} = (\{10^{5},1\},\{10^{5},10\},\{10^{6},1\},\{10^{6},10\},\{10^{7},1\},\{10^{7},10\})\),  from top to bottom, respectively.}
\label{f1}
\end{figure}

In Figure \ref{f1} we show the number of cycles in the last year of inspiral including the contributions from 1PN, 1.5PN and 2PN conservative corrections. We consider six different binary systems which consist of a central Kerr black hole of specific mass \( \bar{M}= M/ M_{\bigodot}= \{10^{5},10^{6},10^{7}\}\) and an orbiting compact object of specific mass \(\bar{m}= m/ M_{\bigodot} =\{1,10\}\). We consider both prograde and retrograde orbits.

\begin{figure}[!hbp]
\centering
\includegraphics{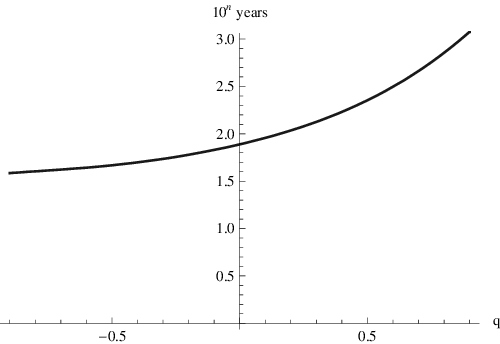}
\caption{This graphic shows the time in years taken for the conservative pieces to contribute one  cycle to the GW phasing for a given binary system as a function of  \(q\). The curve has the same shape for all binary systems. The vertical scale changes according to the binary system under consideration, and is expressed in units of \(10^{n}\) years, where \(n\) depends on the system. Hence if \(\{\bar{M},\bar{m}, n\}\) defines our system, we show the cases: \(\{10^{7},1,4\}, \{10^{7},10,3\}, \{10^{6},1,2\}, \{10^{6},10,1\}, \{10^{5},1,0\}, \{10^{5},10,-1\}\).} 
\label{f2}
\end{figure}

As mentioned before, even though the GW signals are instantaneously  below instrumental and confusion noise, we should be able to detect the signals with high SNR observing them for \(\eta ^{-1}\) cycles, and using matched filtering. But, this requires the template waveform to match the true signal to better than half a cycle over the observation. In Figure \ref{f2} we assess the importance of including the conservative corrections by showing how long all the conservative pieces combined take to contribute one cycle to the GW phasing. We consider both prograde and retrograde orbits for all of the binary systems shown in Figure \ref{f1}. 

\clearpage 

\begin{figure}[!hbp]
\centering
\includegraphics{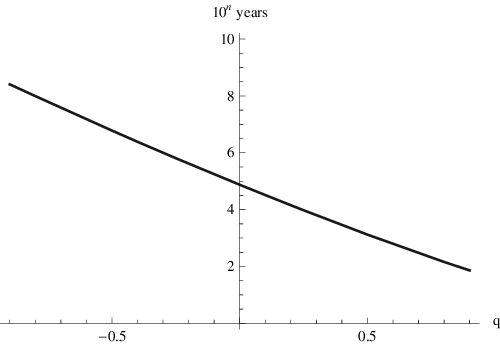}
\caption{This graphic shows the time in years taken for the 2PN conservative piece to contribute one cycle to the GW phasing for a given binary system as a function of \(q\).  The notation is as for Figure~\ref{f2}, and we show the following \(\{\bar{M},\bar{m}, n\}\) cases: \(\{10^{7},1,7\}, \{10^{7},10,6\}, \{10^{6},1,5\}, \{10^{6},10,4\}, \{10^{5},1,3\}, \{10^{5},10,2\}\).}
\label{f3}  
\end{figure}

From Figure \ref{f2} we learn that for a binary system of $10+10^5$ solar masses, the conservative corrections contribute one cycle in about 60 days when \(q=-0.9\). In all other cases it takes longer. The next step is to assess the importance of the highest order conservative term, i.e., the 2PN term. We do this by playing the same game as above, but considering the 2PN term only. This is shown in Figure \ref{f3}.

\clearpage

We learn from Figure \ref{f3} that the 2PN conservative contribution does not have a significant impact on the number of cycles. In fact, referring to the $10+10^5$ binary system considered above, we observe that the 2PN term needs about 141 years to contribute one cycle for a value of \(q=0.9\). In any other case, it takes longer. We also observe that in Figure \ref{f2} the number of years increases as \(q\) increases, whereas Figure \ref{f3} shows the opposite behaviour. This arises because all of the conservative pieces have a positive contribution except the 1.5PN term. Furthermore, this term is the dominant one. Were it not included in the kludge, the number of years taken by the conservative pieces to complete a cycle would be much smaller. However, because the 1.5PN term balances and even overcomes the other positive contributions we get the behaviour portrayed in Figure \ref{f2}. Roughly speaking, an effect will have to be included in a matched filtering template if it contributes of the order of one cycle over the observation. These results suggest that it will be necessary to include conservative corrections in our waveform model, but it may not be necessary to go beyond 2PN order.

\section{A comparison with the self--force formalism for non-spinning black holes}
\label{s3}
The conservative pieces that we have now included in our model, Eq. \eqref{10}, were obtained in the post--Newtonian framework. In the future, we would want to obtain a more accurate solution for the conservative correction by using fits to self--force calculations once these are available. However, it is unlikely that results beyond first order in the mass ratio will be available before LISA flies and we have seen above that second order radiative corrections contribute to the phase evolution at the same level as the first order conservative part of the self--force. Our model includes both corrections, and so we can use it to assess the relative contribution from the first order conservative self--force and the second order dissipative self--force.  To do so, we will use the self--force results recently obtained by Barack and Sago \cite{sago} for a circular orbit around a Schwarzschild black hole. 

The main complication in self--force calculations comes from gauge freedom --- the gauges in which the local and global fields are most easily computed are different. Barack et al. \cite{sago} found a way to circumvent this problem in the non-rotating case by solving both the perturbation equations, and the global retarded field in the Lorenz gauge. In our kludge prescription, by choosing only to modify \(\Omega\) and not \(\mathrm{d} L_z/\mathrm{d}t\) when comparing to the PN model, we have effectively chosen a gauge in which the \(\eta^2\) corrections to the radiative part of the self--force vanish.

\subsection{Assessing self--force calculations using our model}
We can compare self--force results to the kludge results in the same way that we computed the conservative corrections above, namely by comparing asymptotic observables ---  the orbital frequency and its first time derivative. The second order piece of ${\rm d}\Omega/{\rm d}t$ as a function of $\Omega$ depends on both the first order conservative and the second order radiative self--force corrections, and therefore cannot be determined completely from the first order self--force.  We want to understand whether first order accurate self--force calculations will be sufficient for parameter estimation with LISA, by assessing the importance of the second order radiative piece to the orbital phase evolution.

Following \cite{sago}, the orbital frequency of an orbit in the Schwarzschild metric under the influence of the radial self--force, $F_r$, is
\begin{equation} \label{Omega}
\Omega_{\textrm{SF}}={\Omega}_0\left[1-\left(\frac{r(r-3M)}{2M m}\right)F_r\right]\left(1-\frac{m}{\sqrt{r(r-3M)}}\right),
\end{equation}
\noindent where \({\Omega}_0 = \sqrt{M/ r^3}\), and \(r\) stands for the radial coordinate in the self--force framework. 

Barack and Sago \cite{sago} gave a fit to \(F^r\), but the terms they proposed were not motivated by PN expansions. For easier comparison to our results, we can derive an alternative fit to their results of the form  
 \begin{equation}
F^r (r\gg M) \simeq
\frac{m^2}{r^2}
\left[
a_0 + a_1 \frac{M}{r}
+ a_{1.5} \left( \frac{M}{r} \right)^{3/2}
+ a_2 \left( \frac{M}{r} \right)^2
\right],
\label{PNform}
\end{equation}
\noindent where the various coefficients have the values
\begin{eqnarray} \label{a}
&& a_0= 2.01638,\qquad
a_1= -8.98147, \qquad
a_{1.5}= 12.20270,\qquad
a_2= -16.51406.
\end{eqnarray}

\noindent This fit reproduces the numerical data within the numerical  accuracy considered by Barack et al. \cite{sago} in the range \(8M < r <150M\).  Note that the leading term \(F^r \simeq 2.016 m^2/r_0^2\simeq 2m^2/r_0^2\) is consistent with the Keplerian SF which describes the  back--reaction effect that arises from the motion of the black hole around its center of mass. 

To fit the data in the range \(6M < r< 8M\) we can use the expression 
\begin{equation}
F^r(r\ge 6M) \simeq
\frac{m^2}{r^2}
\left[
b_0 + b_1 \frac{M}{r}
+ b_{1.5} \left( \frac{M}{r} \right)^{3/2}
+ b_2 \left( \frac{M}{r} \right)^2
\right],
\label{eq:Fr-xform}
\end{equation}
\noindent where the various coefficients are given by  
\begin{equation}
b_0=2.44194, \quad
b_1=-29.44529, \quad
b_{1.5}=89.73089, \quad
b_2=-99.32447.
\label{co}
\end{equation}
\noindent We can now plug these fits into Eq. \eqref{Omega} to obtain \(\Omega_{\textrm{SF}}\).
\begin{equation}
 \Omega_{\textrm{SF}}={\Omega}_0\left[1+ \eta\left( \lambda_0 + \lambda_1 \frac{M}{r}+ \lambda_{1.5}\left(\frac{M}{r}\right)^{3/2}+ \lambda_2\left(\frac{M}{r}\right)^{2}\right)\right],
\label{om}
\end{equation}

\noindent where the various coefficients have the values

\begin{eqnarray}
6M<r<8M: \, &&\lambda_0= -1.22097, \quad \lambda_1=14.9436, \quad\lambda_{1.5}=-44.8654, \nonumber\\ &&\qquad\qquad\qquad\qquad\lambda_2=35.3815;
\label{coom1}\\
r > 8M: \, &&\hat \lambda_0=-1.00819, \quad \hat \lambda_1=4.49892,\quad\,\, \hat \lambda_{1.5}=-6.10135, \nonumber\\ &&\qquad\qquad\qquad\qquad\hat \lambda_2=4.2826.
\label{sfqco}
\end{eqnarray}

To compare the self--force orbital frequency, Eq. \eqref{om}, to the kludge \(\Omega\), Eq. \eqref{omCC}, we need the coordinate transformation that relates the kludge radial coordinate $p$ to that of the self-force formalism, $r$. We use a transformation of the form
\begin{eqnarray}
\label{cotrans}
p&=& r\Bigg\{1+ \alpha_1 \frac{M}{r}+ \alpha_{1.5}\left(\frac{M}{r}\right)^{3/2}+ \alpha_2\left(\frac{M}{r}\right)^{2} +  \\ \nonumber && \eta\left(\beta_0 +  \beta_1 \frac{M}{r} + \beta_{1.5}\left(\frac{M}{r}\right)^{3/2} + \beta_2 \left(\frac{M}{r}\right)^{2}\right) \Bigg\}.
\end{eqnarray}
\noindent We find the non--vanishing coefficients are given by

\begin{eqnarray}
6M<r<8M: \quad &&\beta_0 =0.89731, \qquad \beta_1 =-8.4929, \qquad \beta_{1.5}=24.25543,  \nonumber\\&&\qquad\qquad\qquad\qquad \beta_2 =-22.2198\; \label{core}\\ 
 r > 8M: \quad &&\beta_0 = 0.75546, \qquad \beta_1 = -1.52978, \qquad \beta_{1.5}= -1.58730, \nonumber\\&&\qquad\qquad\qquad\qquad\beta_2 = -1.15393.
\label{coeff}
\end{eqnarray}

\noindent Using the coordinate transformations, Eqs. \eqref{cotrans}, \eqref{core}, \eqref{coeff}, in Eq.~\eqref{omCC}  recovers Eq.~\eqref{Omega}.

Under the influence of the conservative self--force, the angular momentum of an orbit undergoes a shift
\begin{equation}
L= L_0 \left(1- \frac{r^2}{2 M m}F^r \right),
\label{amp}
\end{equation}
\noindent where \(L_0 = \sqrt{M r^2/(r-3M)}\). The first order radiative self--force is given by Eq.~\eqref{new_Ldot}, but with $p$, the kludge radial coordinate, replaced by $r$, the self--force radial coordinate. As before, the evolution of $\dot{r}$ is best expressed as $\dot{r} = \dot{L_z}/({\rm d}r/{\rm d}L_z)$ with
\begin{eqnarray}
 \frac{1}{M}\frac{{\rm d}L_z}{{\rm d}r} &=& \frac{1}{2(r/M-3)^{3/2}} \left(\frac{r}{M} - 6  +\frac{\eta}{2} \left(\epsilon_{-1} \frac{r}{M} +\epsilon_0 + \epsilon_{0.5} \left(\frac{M}{r}\right)^{1/2}  \right.\right.\nonumber\\ &&\hspace{1.0in} \left.\left.+\epsilon_{1} \left(\frac{M}{r}\right)+ \epsilon_{1.5} \left(\frac{M}{r}\right)^{3/2} +\epsilon_{2} \left(\frac{M}{r}\right)^{2}\right)\right) \label{drdLSF}
\end{eqnarray}
\noindent where the various coefficients can be found to be
\begin{eqnarray}
 6M<r<8M: &&\epsilon_{-1} =-2.44194, \qquad \epsilon_0=-14.7936, \qquad \epsilon_{0.5} = 179.462,\nonumber \\ &&\epsilon_1=-297.973, \qquad\epsilon_{1.5}=-269.193, \qquad \epsilon_2=595.947; \label{sfco}\\
 r > 8M: && \epsilon_{-1} =-2.01638, \qquad \epsilon_0=3.11681, \qquad \epsilon_{0.5} = 24.4054\nonumber\\ &&\epsilon_1=-49.5422, \qquad\epsilon_{1.5}=-36.6081, \qquad \epsilon_2=99.0843. \label{sfco1}
\end{eqnarray}

\noindent By construction, in this gauge, we are missing a contribution from the second order part of $\dot{L_z}$. However, we have already computed a similar expression within the kludge approach, Eq.~\eqref{6}, which includes that contribution. Using the coordinate transformation, Eqs~\eqref{cotrans}--\eqref{coeff}, we can use the kludge result to find the second order piece of $\dot{L_z}$ in the self-force gauge
\begin{eqnarray}
\delta \dot{L_z} = -\frac{32}{5} \eta^2 \left(\frac{M}{r}\right)^{7/2} \Bigg\{l_0 &+&l_{1} \left(\frac{M}{r}\right)\quad+\quad l_{1.5} \left(\frac{M}{r}\right)^{3/2} \quad+\quad l_{2} \left(\frac{M}{r}\right)^{2}\Bigg\} \nonumber \\ \mbox{where}\quad   
6M<r<8M: \quad l_0 &=& -4.0379, \qquad  l_1 = 48.0413, \qquad  l_{1.5} = -140.422,  \nonumber\\&&\qquad\qquad\qquad\qquad l_2 = -57.544; \nonumber\\
r > 8M: \quad l_0 &=& -3.3995, \qquad  l_1 = 20.7749, \qquad  l_{1.5} = -52.1948, \nonumber\\&&\qquad\qquad\qquad\qquad  l_2 = 1.5371. \label{sfKco1}
\end{eqnarray}

The importance of the second order radiative piece of the self--force is readily assessed by computing the difference between an evolution using only Eq.~\eqref{new_Ldot} for $\dot{L_z}$ and one that also includes the second order correction above. We can now compute the number of cycles that the stellar mass compact object (CO) performs before plunge for various binary systems in three separate cases --- 1) ignoring all $\eta^2$ corrections to $\dot{r}$ (i.e., setting $\delta \dot{L_z}$ and the $\epsilon$ coefficients in Eq.~\eqref{drdLSF} to zero); 2) ignoring the second order radiative contribution (i.e., use $\delta \dot{L_z}=0$ and the $\epsilon$ coefficients from Eqs.~\eqref{sfco}--\eqref{sfco1}); 3) including all second order corrections (i.e., as (2) but now with $\delta \dot{L_z}$ from Eq.~\eqref{sfKco1}). The results are given in Table \ref{t3}. From this Table we learn that the second order radiative bit which is missing from the self--force formalism appears to be relatively unimportant for the gravitational wave phasing, since the number of gravitational waveform cycles changes very little as we change the approximation used to compute them. 

\begin{table}[thb]
\centerline{$\begin{array}{lccccccccc}\hline\hline
{M/ M_\odot}     & 0.6 & 1.4 & 10 \\            
\hline\hline
\text{Number of cycles with no 2PN  order corrections}  &                   {133312.6} & {127503.0} & {98642.5}
                 \\ \hline
\text{Number of cycles with no 2PN order radiative corrections}  &          {133311.9} & {127502.1} & {98645.8}
                  \\ \hline
                  \text{Number of cycles with all 2PN order corrections}  & {133312.2} & {127502.6} & {98643.3}
                  \\ \hline
\hline\hline
\end{array}$}
\caption{\protect\footnotesize
In this table we present the number of gravitational waveform cycles that are generated during the last year of inspiral of COs with masses of \(0.6M_\odot, 1.4M_\odot\, \textrm{and}\, 10 M_\odot\) into a $10^6M_{\odot}$ Schwarzschild black hole.  We consider three different approximations, as described in the text.}
\label{t3}
\end{table}

\section{Noise induced parameter errors}
\label{s4}
We now want to look at the parameter estimation accuracy which LISA observations are likely to achieve, for which we need waveforms in addition to the orbital phase evolution. We use the standard numerical kludge approach described in Babak et al. \cite{kludge} and refer the reader to \cite{mtw} for background. The waveform is generated by using a flat--space quadrupole wave generation formula applied to the trajectory of the inspiralling object in Boyer-Lindquist coordinates, which are identified with spherical-polar coordinates in the flat-space. In the transverse--traceless (TT) gauge, the metric perturbation \(h_{i j}(t)\) is given by 

\begin{equation}
h_{i j}(t) = \frac{2}{D}\left(P_{ik}P_{jl}- \frac{1}{2}P_{ij}P_{kl}\right)\ddot  I^{kl},
\label{13}
\end{equation}

\noindent where \(D\) is the distance to the source, \(\eta_{ij}\) is the flat metric, the  projection operator is given by \(P_{ij} = \eta_{ij} - \hat n_{i}\hat n_{j}\), \(\hat n_{i}\) is the unit vector in the direction of propagation, and \(\ddot  I^{kl}\) is the second time derivative of the inertia tensor \cite{mtw}. In the EMRI framework it takes the form \(  I^{kl}= m r^i(t)r^j(t)\), where \(r^i(t)\) represents the position vector of the compact object with respect to the MBH in the pseudo--flat space. 

\subsection{Implementation of LISA's response function}

Following \cite{cutlerold}, the LISA response may be written as
\begin{equation}
h_{\alpha}(t)= \frac{\sqrt{3}}{2 D} \Big[F_\alpha ^ +(t)A^ +(t) + F_\alpha ^ \times(t) A^ \times(t)\Big],
\label{14}
\end{equation}
\noindent where $\alpha = I,II$ refers to the two independent Michelson-like detectors that constitute the LISA response at low frequencies. The functions \(A^ {+\, , \times} (t)\) are the polarization coefficients given by
\begin{equation}
A ^ + = -a_+ [1+ (\hat{a} \cdot \hat n)^2], \qquad A^ \times = 2 a_\times (\hat a \cdot \hat n),
\label{15}
\end{equation}
\noindent where \(\hat a\) is a unit vector along the SMBH's spin direction, and \(a_+,a_\times\) are given by 
\begin{displaymath}
a_+= \frac{1}{2}\left( \ddot  I^{11}- \ddot  I^{22}\right), \qquad a_\times= \ddot  I^{12},
\end{displaymath}
\noindent The antenna pattern functions \(F_\alpha ^{+ \times}\) are given by
\begin{eqnarray}  
F^{+}_{I} &=&  \frac{1}{2}(1+\cos^2\theta)\cos(2\phi)\cos(2\psi)
-\cos\theta\sin(2\phi)\sin(2\psi), \nonumber\\
F^{\times}_{I} &=& \frac{1}{2}(1+\cos^2\theta)\cos(2\phi)\sin(2\psi)
+\cos\theta\sin(2\phi)\cos(2\psi),
\label{16} \\
F^{+}_{II} &=&  \frac{1}{2}(1+\cos^2\theta)\sin(2\phi)\cos(2\psi)
+\cos\theta\cos(2\phi)\sin(2\psi), \nonumber\\
F^{\times}_{II} &=& \frac{1}{2}(1+\cos^2\theta)\sin(2\phi)\sin(2\psi)
-\cos\theta\cos(2\phi)\cos(2\psi).
\label{17}
\end{eqnarray}  

\noindent The various angles in the previous expressions represent the source's sky location in a detector based coordinate system, (\(\theta,\phi\)), and the polarization angle of the wavefront, \(\psi\). These can be re--written in a fixed, ecliptic--based coordinate system. If we denote the source co--latitude and azimuth angles and the direction of \(\hat a\) in this fixed coordinate system by (\(\theta_S,\phi_S\)) and (\(\theta_K,\phi_K\)) respectively, then 
\begin{eqnarray} 
\cos\theta(t) &=& \frac{1}{2}\cos\theta_S-\frac{\sqrt{3}}{2}\sin\theta_S
\cos[\bar\phi_0+2\pi(t/T)-\phi_S], \nonumber\\
\phi(t) &=& \bar\alpha_0+2\pi(t/T)+ \tan^{-1}\Bigg\{
\frac{\sqrt{3}\cos\theta_S+\sin\theta_S\cos[\bar\phi_0+2\pi(t/T)-\phi_S]}
{2\sin\theta_S\sin[\bar\phi_0+2\pi(t/T)-\phi_S]}\Bigg\},\nonumber\\
\tan\psi & = & \Bigg\{\frac{1}{2}\cos\theta_K-\frac{\sqrt{3}}{2}
\sin\theta_K \cos[\bar\phi_0+2\pi(t/T)-\phi_K] \nonumber \\ &-& \cos\theta(t)\left[
\cos\theta_K\cos\theta_S+\sin\theta_K\sin\theta_S\cos(\phi_K-\phi_S)\right] \Bigg\}\Big /  
 \nonumber\\ && \Bigg\{
\frac{1}{2}\sin\theta_K\sin\theta_S\sin(\phi_K-\phi_S)
-\frac{\sqrt{3}}{2}\cos(\bar\phi_0+2\pi t/T)\nonumber\\ &&\{\cos\theta_K\sin\theta_S\sin\phi_S
-\cos\theta_S\sin\theta_K\sin\phi_K\} \nonumber\\ &-&
\frac{\sqrt{3}}{2}\sin(\bar\phi_0+2\pi t/T) \left(\cos\theta_S\sin\theta_K\cos\phi_K
-\cos\theta_K\sin\theta_S\cos\phi_S\right)\Bigg\},\label{18}
\end{eqnarray}
\noindent where \( \bar\phi_0, \bar\alpha_0\) are constant angles which represent the orbital and rotational phase of the detector at \(t=0\). We will set both of these to zero in our analysis. Additionally, \(T\) is the orbital period, which is 1 year. Barack and Cutler \cite{cutler} write these expressions in terms of  \(\theta_L, \phi_L\) which specify the direction of the compact object's orbital angular momentum in the ecliptic--based system. In this case the orbits under consideration are circular and equatorial, so the angular momentum vector of the orbiting body does not precess about the SMBH's spin \(\hat{a}\) and \(\theta_K =\theta_L , \phi_K = \phi_L\). 

The last ingredient in the detector response is the Doppler phase modulation. If \(\Phi(t)\) denotes the phase of the waveform, the inclusion of the Doppler modulation shifts the phase as follows \cite{cutler}
\begin{equation}
\Phi(t)\to \Phi(t)+ 2 \frac{\mathrm{d}\phi}{\mathrm{d}t} R \sin\theta_S \cos[2\pi(t/T)-\phi_S],
\label{20}
\end{equation}
\noindent where  \(R= 1 \textrm{AU/c}= 499.00478 \textrm{s}\) and \(\mathrm{d}\phi/\mathrm{d}t\) is the azimuthal velocity of the orbit, cf. \eqref{1.3}.

\subsection{Signal analysis}
In this section we will briefly review the basics of signal analysis. We can think of a GW detector as a linear system whose input is a GW we want to detect and whose output is  a time series. This time series is  a combination of both the instrumental noise and a true GW signal. For LISA, the output of each of the two equivalent two arm Michelson detectors can be represented as
\begin{equation}
s_{\alpha}(t) = h_{\alpha}(t) + n_{\alpha}(t), \qquad \alpha = \textrm{I, \,II}.
\label{21}
\end{equation}
\noindent The detection problem is to identify \( h_{\alpha}(t)\) in the presence of \( n_{\alpha}(t)\). Making the usual assumption that each Fourier component  of the noise, \({\tilde n_{\alpha}}(f)\), is Gaussian distributed, and uncorrelated with other Fourier components (i.e., the noise is stationary), the ensemble average of the Fourier components of the noise have the property 
\begin{equation}
\label{22}
\langle {\tilde n_\alpha}(f) \, {\tilde n_\beta}(f^\prime)^* \rangle =\frac{1}{2}
\delta(f - f^\prime) S_n(f) \delta_{\alpha \beta}.
\end{equation}
\noindent This relation defines the one-sided spectral density of the instrumental noise, \(S_n(f)\). For LISA, the spectral density is the same in the two detectors.

The problem we have to solve is how to estimate the parameters of the system. That is, if the expected waveform \(h(t;\theta)\) depends on parameters \(\theta=\{\theta_1,...,\theta_N\}\), we want to reconstruct the most probable value of the parameters of the source and compute their respective errors. To effectively address this problem we need to calculate the probability of the parameters  given the data, i.e., the posterior probability. We may define a natural inner product on the vector space of signals, which for any two signals \(p_\alpha (t), q_\alpha(t)\), takes the form
\begin{equation}\label{23}
\left( {\bf p} \,|\, {\bf q} \right)
\equiv 2\sum_{\alpha} \int_0^{\infty}\left[ \tilde p_\alpha^*(f)
\tilde q_\alpha(f) + \tilde p_\alpha(f) \tilde q_\alpha^*(f)\right]
/S_n(f)\,\mathrm{d}f.
\end{equation}
\noindent The probability distribution function for the noise \(n(t)\) is given by 
\begin{equation}
\label{24}
p({\bf n}_0) = N \textrm{exp}\left(- \frac{\left( {\bf n}_0\, |\,{\bf n}_0 \right)}{2}\right),
\end{equation}
\noindent where \(N\) is a normalization factor. Assuming we have made a detection, the output has the form \({\bf s}(t)= {\bf h}(t;\theta_{\textrm{true}}) + {\bf n}_0(t)\), where \(n_0(t)\) is the specific realization of the noise and \(\theta_{\textrm{true}}\) is the unknown true value of the parameters. The probability of measuring this output signal is then
\begin{equation}
\label{25}
p\left( {\bf h}(\theta_{\textrm{true}}) \,|\, {\bf s} \right) = N\,p_0(\theta_{\textrm{true}}) \textrm{exp}\left( \left( {\bf h}(\theta_{\textrm{true}})\, |\, {\bf s} \right)- \frac{1}{2}\left( {\bf h}(\theta_{\textrm{true}})\, |\, {\bf h}(\theta_{\textrm{true}}) \right)\right),
\end{equation} 
\noindent where \(p_0(\theta_{\textrm{true}})\) is a prior probability \cite{maggiore}.  This probability distribution function encodes all the information we can extract from the data stream. We want to determine the most probable value of the parameters and estimate the likely errors in these estimates. The maximum likelihood estimators of the parameters are those that minimize the product \(\left( {\bf s-h}\,|\, {\bf s-h} \right) \). This choice, which also provides the highest  SNR in a matched filtering search, is given by
\begin{equation}
\frac{S}{N}[h(\theta^i)]= \frac{\left( {\bf s}\,|\, {\bf h} \right)}{\sqrt{\left( {\bf h}\,|\, {\bf h} \right)}}.
\label{26}
\end{equation}
\noindent If the SNR is large, the parameter estimation error is likely to be small. Ignoring the prior, we can expand Eq.~\eqref{25} about the peak, $\hat{\theta}$ by setting $\theta^i= \hat{\theta^i} + \Delta \theta^i$ and find 
\begin{equation}
\label{27}
p(\Delta\theta \, |\,s)=\,{\cal N} \, e^{-\frac{1}{2}\Gamma_{ij}\Delta \theta^i
\Delta \theta^j}, \qquad \Gamma_{ij} \equiv \bigg( \frac{\partial {\bf h}}{ \partial \lambda^i}\, \bigg| \,
\frac{\partial {\bf h}}{ \partial \lambda^j }\bigg)_{|\theta=\hat{\theta}}.
\end{equation}
\noindent $\Gamma_{ij}$ is the Fisher Information Matrix. For large SNR, the expectation value of the errors \(\Delta \theta^i\) is given by
\begin{equation}
\label{29}
\left< {\Delta \theta^i} {\Delta \theta^j}
 \right>  = (\Gamma^{-1})^{ij} + {\cal O}({\rm SNR})^{-1} .
\end{equation}
\noindent In our analysis, we employ a simplified definition of the inner product, Eq. \eqref{23}. For white noise, i.e., \(S_n(f)= \textrm{const.}\), it would take the simple form  \(2 S_n^{-1}\sum_{\alpha}\int_{-\infty}^{\infty} \, p_\alpha(t) q_\alpha(t) dt\), by Parseval's theorem \cite{cutler}. If we define the ``noise--weighted'' waveform
\begin{equation}\label{30}
\hat h_{\alpha}(t) \equiv   \frac{h_{\alpha}(t)}{\sqrt{S_h\bigl(f(t)\bigr)}}, \qquad f(t) = \frac{1}{\pi}\frac{\mathrm{d}\phi}{\mathrm{d}t},
\end{equation}
\noindent we can rewrite the Fisher matrix approximately as 
\begin{equation}\label{31}
\Gamma_{ab} = 2\sum_{\alpha}\int_0^T{\partial_a \hat h_{\alpha}(t) \partial_b \hat h_{\alpha}(t) \mathrm{d}t} \, .
\end{equation}
\noindent This approach was also used by Barack and Cutler \cite{cutler}. 

\subsection{Noise model}
The function \(S_h\bigl(f\bigr)\) is the total LISA noise, which has three components: instrumental noise, confusion noise from short--period galactic binaries, and confusion noise from extragalactic binaries. We use the same prescription as in Barack and Cutler~\cite{cutler}, namely
\begin{eqnarray}
S_h\bigl(f(t)\bigr)& =& {\rm min}\big\{S_h^{\rm inst}(f)/\exp(-\kappa T_{\rm mission}^{-1} \mathrm{d}N/\mathrm{d}f) + S_h^{\rm ex gal}(f) ,\nonumber\\ &&  S_h^{\rm inst}(f) + S_h^{\rm gal}(f)+  S_h^{\rm ex gal}(f)\big\}\, ,
\label{32}
\end{eqnarray}
\noindent where the various components have the following analytic forms \cite{cutler}
\begin{eqnarray}
S^{\rm inst}_h(f) &=& 9.18 \times 10^{-52}f^{-4} + 1.59 \times 10^{-41}
+ 9.18 \times 10^{-38}f^{2}\;\; {\rm Hz}^{-1}, \nonumber \\ 
S^{\rm gal}_h(f) &=& 2.1\times10^{-45}\,\left(\frac{f}{1{\rm Hz}}\right)^{-7/3} \; {\rm Hz}^{-1}, \nonumber \\ 
S_h^{\rm ex.\ gal} &=&
4.2 \times 10^{-47} \left(\frac{f}{1{\rm Hz}}\right)^{-7/3}\; {\rm Hz}^{-1}.
\label{33}
\end{eqnarray}
\noindent Here \(\mathrm{d}N/\mathrm{d}f\) is the number density of galactic white dwarf binaries per unit GW frequency, and \(\kappa\) is the average number of frequency bins that are lost when each galactic binary is fitted out. We use
\begin{equation}\label{34}
\frac{\mathrm{d}N}{\mathrm{d}f} = 2\times10^{-3}\,{\rm Hz}^{-1}\left(\frac{1\,{\rm Hz}}{f}\right)^{11/3}, \qquad \kappa T_{\rm mission}^{-1} = 1.5/{\rm yr},
\end{equation}
\noindent with \( T_{\rm mission} = 3\) yr, and \(\kappa =4.5\).

\subsection{Parameter estimation error results}
To estimate errors using the inverse Fisher Matrix we used fixed values of the intrinsic parameters of the source and carried out a Monte Carlo simulation over values of the extrinsic parameters. The error estimate is SNR dependent, but we quote results at fixed SNR$=30$. We first compute the Fisher Matrix for a source at $D=1$Gpc, and the corresponding SNR from the expression
\begin{equation}\label{35}
{\rm SNR}^2= 2\sum_{\alpha=I,II}\int_{t_{\rm init}}^{t_{\rm LSO}}
\hat h_{\alpha}^2(t)dt.
\end{equation}
\noindent We then multiply the errors from the inverse Fisher Matrix by (SNR$/30$) to normalise to SNR$=30$. We considered one year observations and chose the initial radial coordinate, $p_0$, such that the compact object would reach the last stable orbit after one year of inspiral. The Fisher Matrix has ten dimensions. Four of these are intrinsic parameters, namely \(\ln m, \ln M, q, p_0\). The other six are extrinsic or phase parameters. We summarize the physical meaning of the parameters in Table \ref{tableparams}. 

\begin{figure*}[ht]
\centerline{
\includegraphics[height=0.5\textwidth,angle=270,  clip]{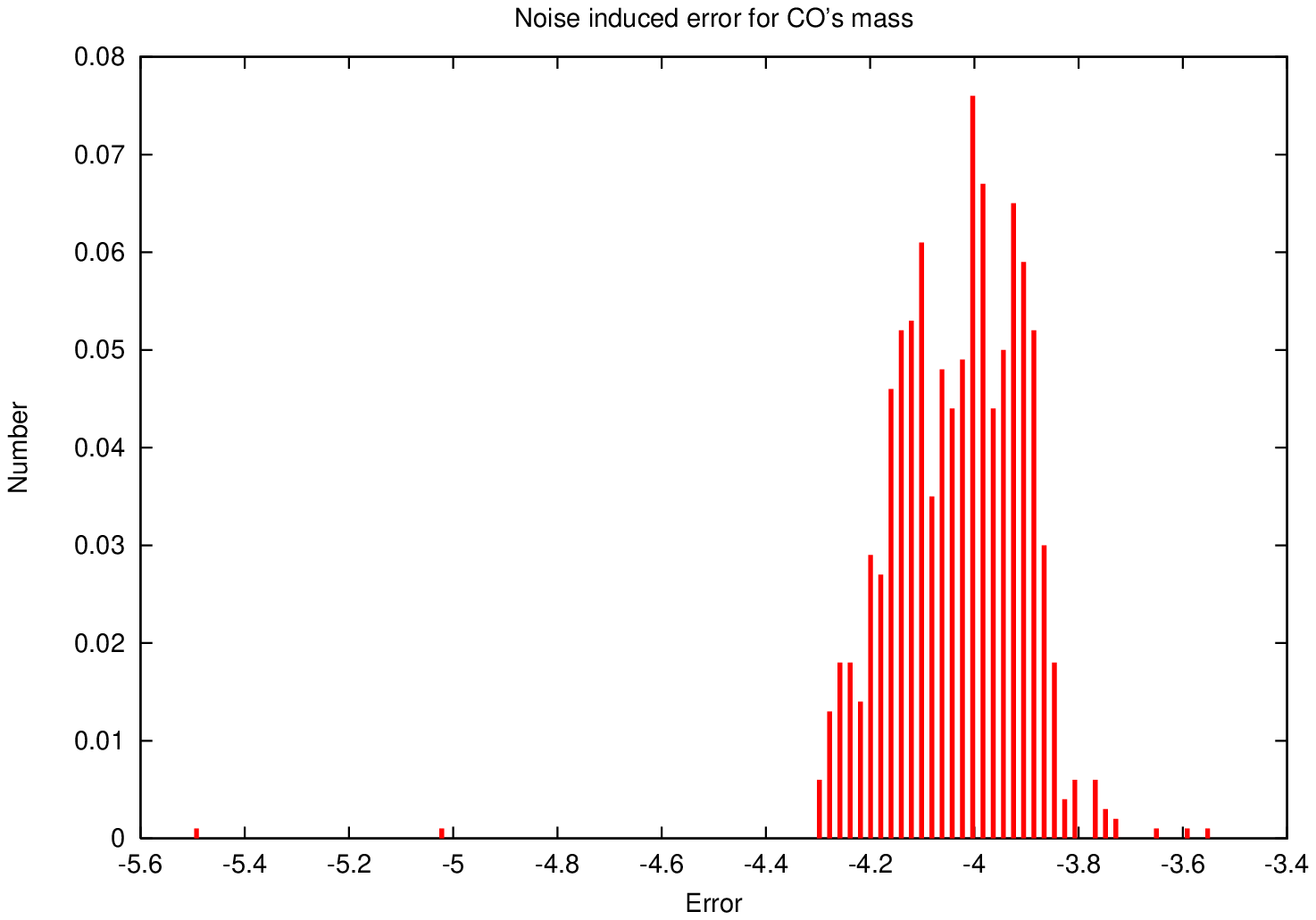}
\includegraphics[height=0.5\textwidth, angle=270,clip]{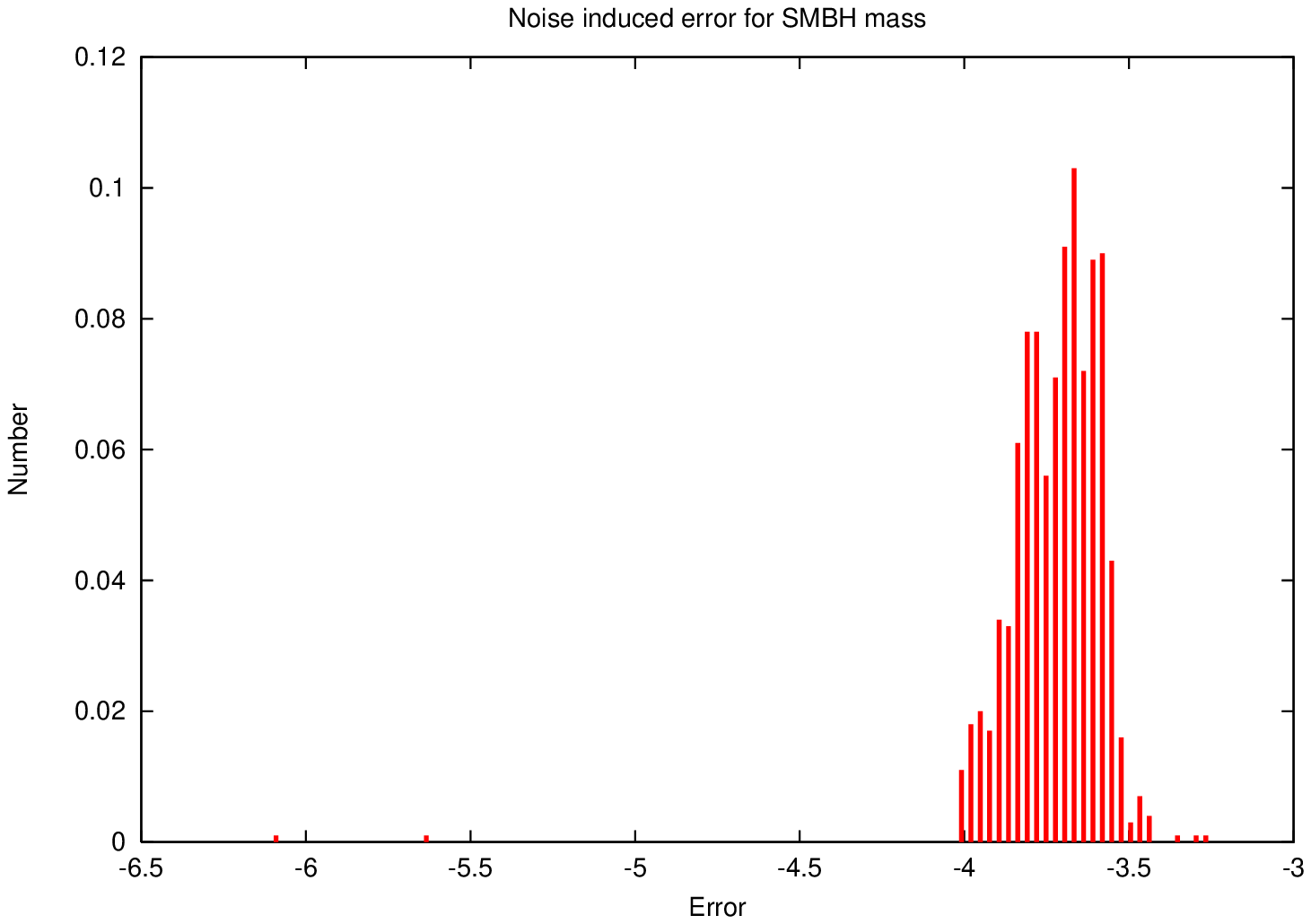}
}
\centerline{
\includegraphics[height=0.5\textwidth, angle=270, clip]{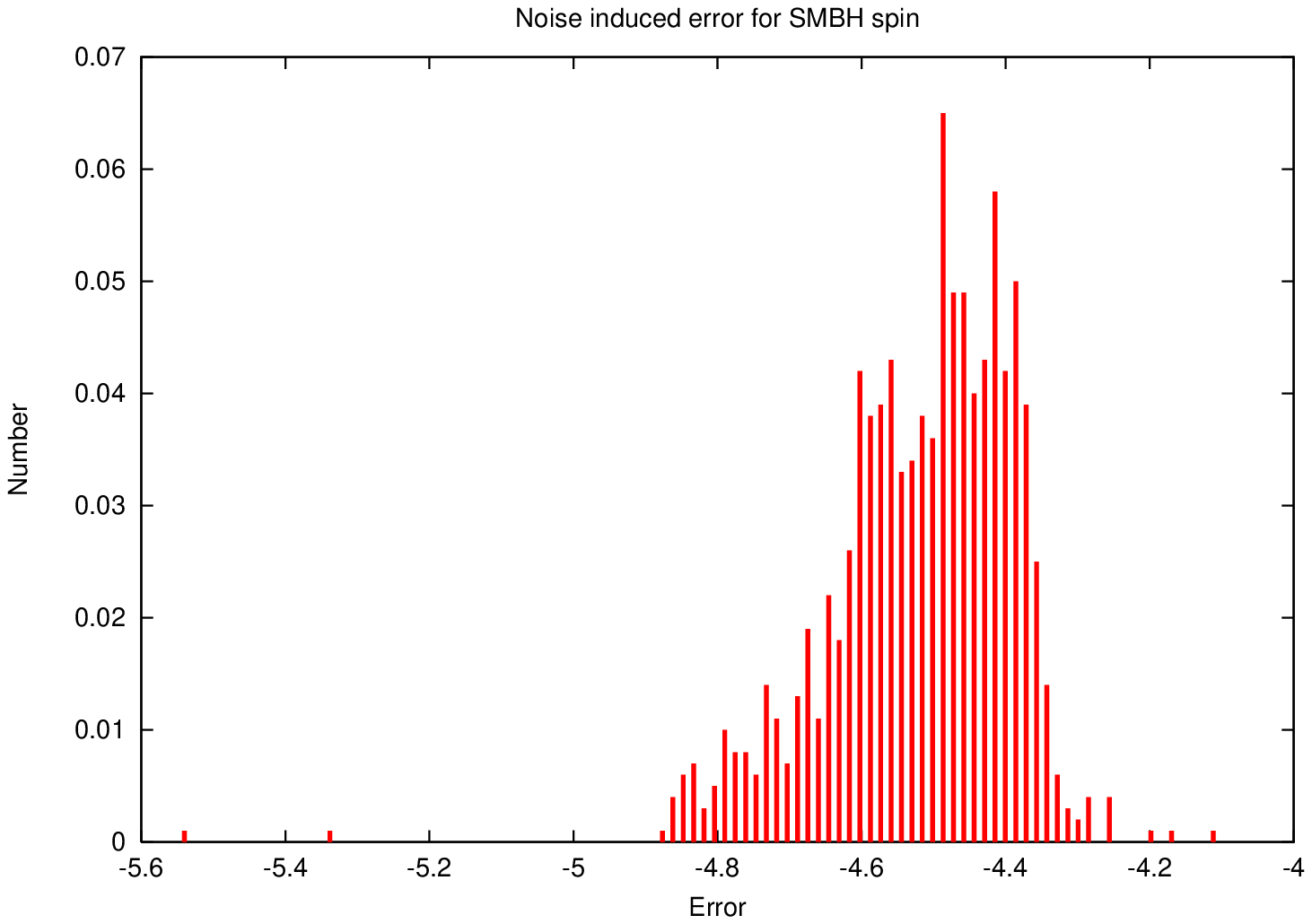}
\includegraphics[height=0.5\textwidth,angle=270, clip]{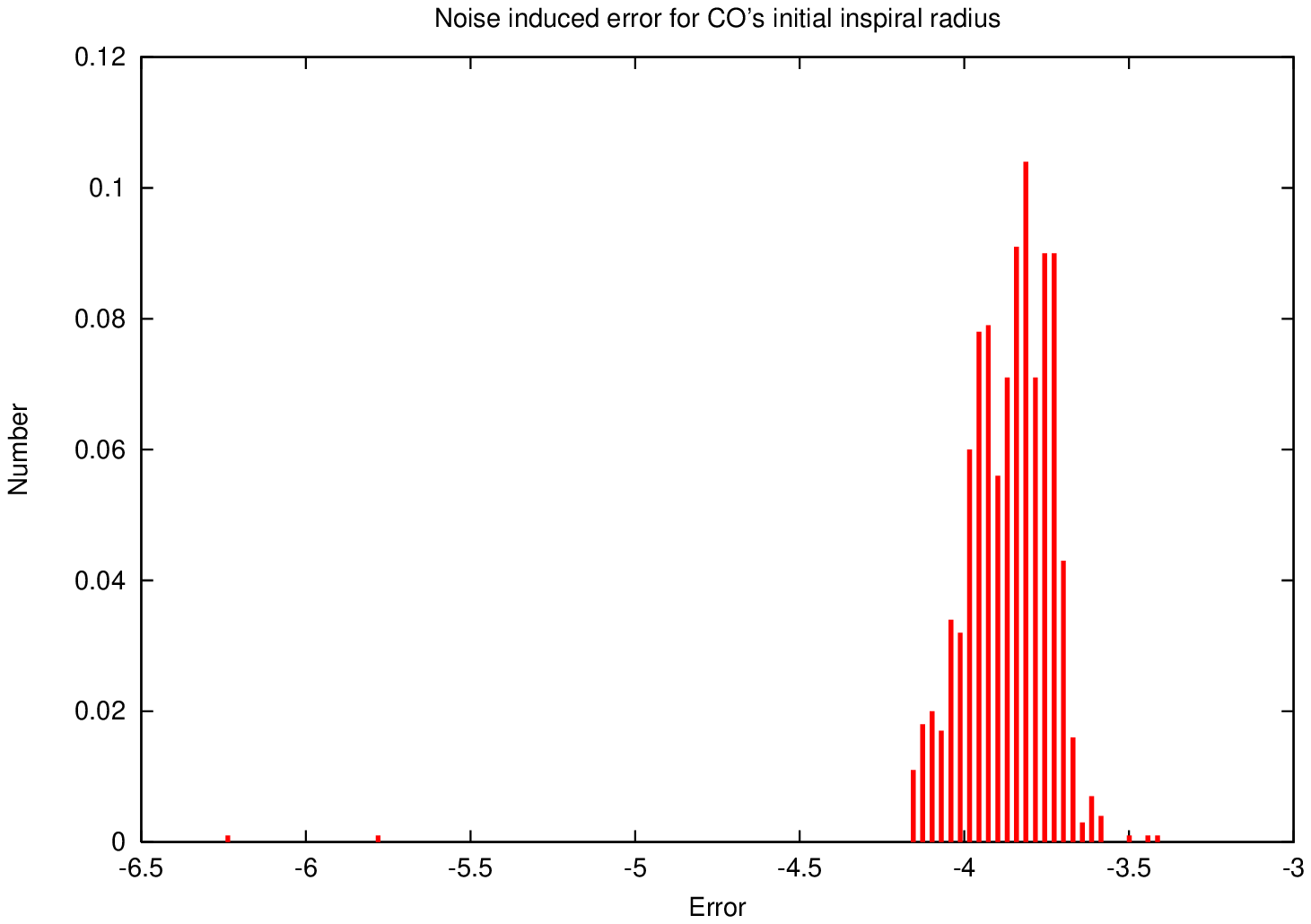}
}
\caption{Distributions of errors in the intrinsic parameters as computed from the Monte Carlo simulations of the inverse Fisher Matrix. We show the system with $m=10M_{\odot}$, $M=10^6M_{\odot}$, $q=0.9$ and show the errors in $\ln(m)$, $\ln(M)$, $\ln(q)$ and $\ln(p_0)$ respectively.}
\label{mcfmdis}
\end{figure*}

For these Monte Carlo simulations, we evolved the orbit via Eq.~\eqref{6}, i.e., evolving $L_z$ rather than $p$, included the full geodesic frequency in Eq.~\eqref{7.1} and included the higher order Teukolsky fits in Eq.~\eqref{new_Ldot}. In this way, we included lower order effects as accurately as possible as discussed earlier. The results of the Monte Carlo simulations are summarized in Tables \ref{tabFMErrBH}, \ref{tabFMErrNS} and \ref{tabFMErrWD}. We considered ``typical'' systems with $M=10^6M_{\odot}$, $q=0.9$ and three different values of $m=0.6M_{\odot}, 1.4M_{\odot}, 10M_{\odot}$ to represent inspirals of white dwarfs, neutron stars and black holes, respectively. For comparison to the self force we also considered a case with $q=0$, for which we also ignored $q$ as a parameter in the Fisher Matrix. The tables list the mean, standard deviation, median and lower and upper quartiles of the distribution of Fisher Matrix errors computed in the Monte Carlo simulation. In Figure~\ref{mcfmdis} we show sample histograms of the Fisher Matrix errors in the intrinsic parameters computed from the Monte Carlo simulation for the $m=10M_{\odot}$ system. 

For the $m=10M_{\odot}$ case we also carried out Monte Carlo simulations with (a) $\dot{L_z}$ truncated at 2PN order, which did not significantly affect the results for the $q=0$ case, but made the $q=0.9$ errors greater by about an order of magnitude; and (b) $p$ evolved directly via the 2PN expression, Eq.~\eqref{7}, rather than evolving $L_z$, which made the errors for the $q=0$ case about an order of magnitude greater, but which did not significantly change the $q=0.9$ results.

\setlength{\tabcolsep}{7pt}
\setlength{\extrarowheight}{1.5pt}
\begin{table}[thb]
\centerline{$\begin{array}{c|l}\hline\hline
 \ln m      & \text {mass of CO}   \\
 \ln M         & \text {mass of SMBH}\\
 q         & \text{magnitude of (specific) spin
    angular momentum of SMBH} \\
p_0           & \text{Initial radius of CO orbit} \\
\phi_0 &  \text{Initial phase of CO orbit}      \\
\theta_S  &  \text{source sky colatitude in an ecliptic--based system }  \\
\phi_S   & \text{source sky azimuth in an ecliptic--based system}  \\
 \theta_K   & \text{direction  of SMBH spin (colatitude)}  \\
\phi_K       &  \text{direction of SMBH spin (azimuth)}  \\
\ln D          & \text{distance to  source}\\
\hline\hline
\end{array}$}
\caption{\protect\footnotesize
This table describes the meaning of the parameters used in our model. The angles ($\theta_S$,\,$\phi_S$) and ($\theta_K$,\,$\phi_K$) are defined in a fixed ecliptic--based coordinate system.}
\label{tableparams}
\end{table}

\begin{table}[thb]
\begin{tabular}{|c|c|c|c|c|c|c|c|c|c|c|c|}
\hline\multicolumn{2}{|c|}{}&\multicolumn{10}{c|}{Statistics of distribution of \(\log_{10}(\Delta X)\) for error, \(\Delta X\), in parameter \(X =\)}\\\cline{3-12}
\multicolumn{2}{|c|}{Model}&$\ln m$&$\ln M$&$q$&$p_0$&$\phi_0$&$\theta_S$&$\phi_S$&$\theta_K$&$\phi_K$ &$\ln D$\\\hline
&Mean&-4.04&-3.73&-4.53&-3.88&-0.73&-1.63&-1.74&-0.93&-0.79&-1.58\\\cline{2-12}
&St. Dev.&0.121&0.125&0.129&0.125&0.864&0.163&0.205&0.750&0.773&0.391\\\cline{2-12}
Kludge&L. Qt.&-4.14&-3.82&-4.60&-3.96&-1.33&-1.75&-1.90&-1.51&-1.38&-1.85\\\cline{2-12}
(q=0.9)&Med.&-4.04&-3.72&-4.52&-3.87&-1.07&-1.62&-1.76&-1.16&-1.02&-1.72\\\cline{2-12}
&U. Qt.&-3.94&-3.63&-4.43&-3.77&-0.34&-1.50&-1.65&-0.53&-0.37&-1.41\\\hline
&Mean&-4.37&-4.79&N/A&-4.96&-0.70&-1.57&-1.64&-0.90&-0.76&-1.26\\\cline{2-12}
&St. Dev.&0.128&0.124&N/A&0.124&0.904&0.136&0.204&0.770&0.783&0.407\\\cline{2-12}
Self Force&L. Qt.&-4.44&-4.85&N/A&-5.02&-1.33&-1.66&-1.78&-1.48&-1.35&-1.54\\\cline{2-12}
(q=0)&Med.&-4.34&-4.77&N/A&-4.94&-1.05&-1.54&-1.66&-1.13&-0.95&-1.40\\\cline{2-12}
&U. Qt.&-4.27&-4.69&N/A&-4.86&-0.36&-1.45&-1.55&-0.54&-0.41&-1.12\\\hline
\end{tabular}
\caption{Summary of Monte Carlo over Fisher Matrix errors for black hole systems ($m=10M_{\odot}$). We show the mean, standard deviation, median and quartiles of the distribution of the logarithm to base ten of the error in each parameter. Results are given both for the kludge model (with conservative corrections to 1.5PN order) and for the self-force model, as indicated. The angles $\bar\phi_0$ and $\bar\alpha_0$, specifying LISA's position and orientation at $t=0$, are set to zero. }
\label{tabFMErrBH}
\end{table}

\begin{table}
\begin{tabular}{|c|c|c|c|c|c|c|c|c|c|c|c|}
\hline\multicolumn{2}{|c|}{}&\multicolumn{10}{c|}{Statistics of distribution of \(\log_{10}(\Delta X)\) for error, \(\Delta X\), in parameter \(X =\)}\\\cline{3-12}
\multicolumn{2}{|c|}{Model}&$\ln m$&$\ln M$&$q$&$p_0$&$\phi_0$&$\theta_S$&$\phi_S$&$\theta_K$&$\phi_K$ &$\ln D$\\\hline
&Mean&-4.60&-3.94&-4.59&-4.06&-0.75&-1.71&-1.88&-0.96&-0.82&-0.92\\\cline{2-12}
&St. Dev.&0.109&0.118&0.122&0.118&0.882&0.199&0.212&0.766&0.797&0.399\\\cline{2-12}
Kludge&L. Qt.&-4.70&-4.02&-4.65&-4.14&-1.36&-1.87&-2.03&-1.56&-1.40&-1.19\\\cline{2-12}
(q=0.9)&Med.&-4.61&-3.94&-4.58&-4.05&-1.09&-1.71&-1.89&-1.19&-1.03&-1.06\\\cline{2-12}
&U. Qt.&-4.52&-3.85&-4.49&-3.97&-0.42&-1.54&-1.77&-0.60&-0.45&-0.79\\\hline
&Mean&-4.12&-4.87&N/A&-5.06&-0.76&-1.63&-1.71&-0.94&-0.80&-0.62\\\cline{2-12}
&St. Dev.&0.136&0.130&N/A&0.131&0.852&0.155&0.217&0.731&0.767&0.379\\\cline{2-12}
Self Force&L. Qt.&-4.19&-4.95&N/A&-5.13&-1.37&-1.74&-1.86&-1.52&-1.40&-0.88\\\cline{2-12}
(q=0)&Med.&-4.10&-4.86&N/A&-5.04&-1.08&-1.61&-1.73&-1.17&-1.01&-0.76\\\cline{2-12}
&U. Qt.&-4.02&-4.78&N/A&-4.96&-0.41&-1.49&-1.62&-0.58&-0.40&-0.48\\\hline
\end{tabular}
\caption{As Table~\ref{tabFMErrBH}, but for neutron star inspirals ($m=1.4M_{\odot}$).}
\label{tabFMErrNS}
\end{table}

\begin{table}
\begin{tabular}{|c|c|c|c|c|c|c|c|c|c|c|c|}
\hline\multicolumn{2}{|c|}{}&\multicolumn{10}{c|}{Statistics of distribution of \(\log_{10}(\Delta X)\) for error, \(\Delta X\), in parameter \(X =\)}\\\cline{3-12}
\multicolumn{2}{|c|}{Model}&$\ln m$&$\ln M$&$q$&$p_0$&$\phi_0$&$\theta_S$&$\phi_S$&$\theta_K$&$\phi_K$ &$\ln D$\\\hline
&Mean&-4.78&-3.93&-4.52&-4.03&-0.73&-1.74&-1.88&-0.94&-0.80&-0.57\\\cline{2-12}
&St. Dev.&0.119&0.182&0.195&0.182&0.892&0.211&0.228&0.781&0.813&0.410\\\cline{2-12}
Kludge&L. Qt.&-4.82&-4.00&-4.57&-4.10&-1.36&-1.89&-2.04&-1.56&-1.43&-0.85\\\cline{2-12}
(q=0.9)&Med.&-4.76&-3.91&-4.49&-4.01&-1.07&-1.74&-1.90&-1.15&-1.01&-0.71\\\cline{2-12}
&U. Qt.&-4.70&-3.83&-4.41&-3.93&-0.36&-1.56&-1.77&-0.55&-0.41&-0.43\\\hline
&Mean&-3.99&-4.90&N/A&-5.08&-0.70&-1.64&-1.74&-0.90&-0.77&-0.26\\\cline{2-12}
&St. Dev.&0.139&0.132&N/A&0.133&0.928&0.157&0.196&0.799&0.819&0.424\\\cline{2-12}
Self Force&L. Qt.&-4.05&-4.98&N/A&-5.16&-1.37&-1.76&-1.89&-1.52&-1.39&-0.56\\\cline{2-12}
(q=0)&Med.&-3.96&-4.88&N/A&-5.06&-1.05&-1.63&-1.74&-1.13&-0.98&-0.42\\\cline{2-12}
&U. Qt.&-3.88&-4.80&N/A&-4.98&-0.29&-1.51&-1.64&-0.48&-0.36&-0.11\\\hline
\end{tabular}
\caption{As Table~\ref{tabFMErrBH}, but for white dwarf inspirals ($m=0.6M_{\odot}$).}
\label{tabFMErrWD}
\end{table}

\begin{figure*}[htp]
\centerline{
\includegraphics[width=0.5\textwidth,  clip]{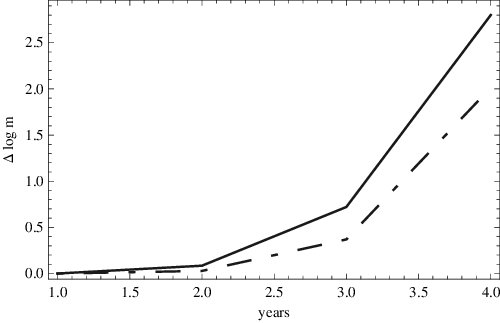}
\includegraphics[width=0.5\textwidth, clip]{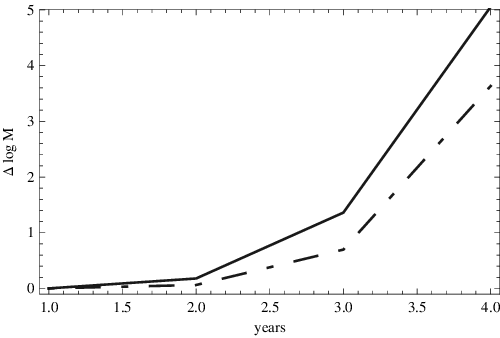}
}
\centerline{
\includegraphics[width=0.5\textwidth,  clip]{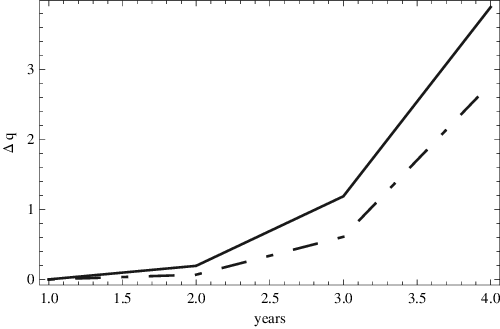}
\includegraphics[width=0.5\textwidth, clip]{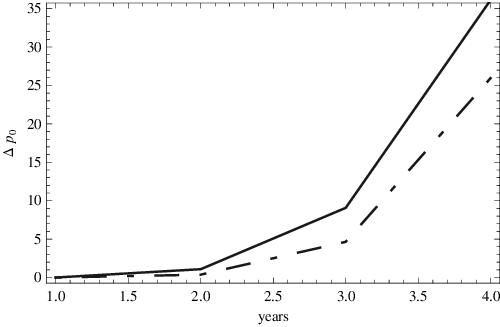}
}
\caption{ We show plots of noise--induced errors in a one year observation as a function of time remaining until plunge at the start of the observation. Two cases are illustrated, namely (1) the distance to the source is normalized by means of weighting the noise errors by a factor of \(\textrm{SNR}/30\) (solid line),  (2)  the distance to the source is fixed as in (1) but only for the last year before plunge (dot-dash line). The plots correspond to a \( 10M_{\odot}\) CO orbiting around a  \( 10^6 M_{\odot}\) SMBH with parameter spin \(q=0.9\). We have set the various extrinsic  parameters as follows: \(\Phi_0 =0\), \(\theta_S= \pi/4\), \(\phi_S=0\), \(\theta_K=\pi/8\), \(\phi_K=0\). These four plots show the errors in the intrinsic parameters.}
\label{fn1}
\end{figure*}

\begin{figure*}[htp]
\centerline{
\includegraphics[width=0.5\textwidth,  clip]{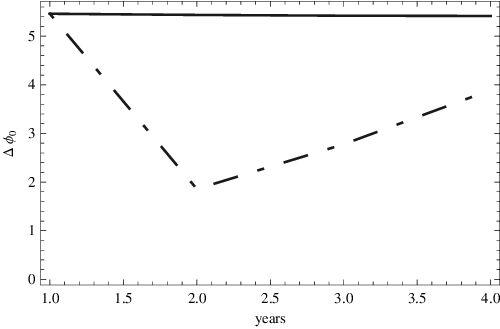}
\includegraphics[width=0.5\textwidth, clip]{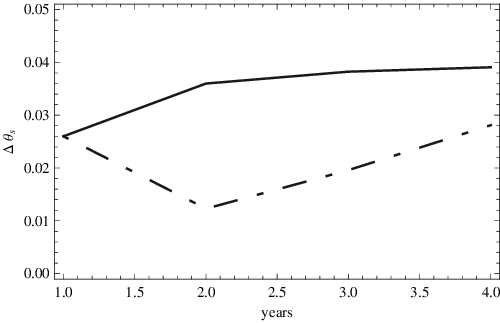}
}
\centerline{
\includegraphics[width=0.5\textwidth, clip]{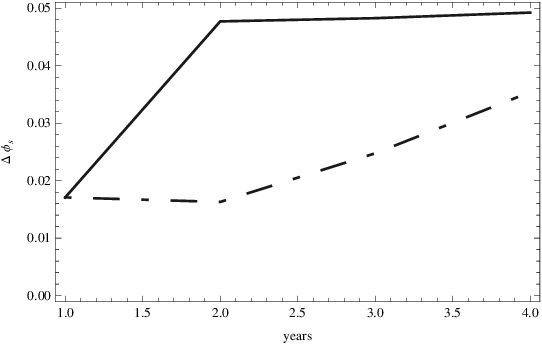}
\includegraphics[width=0.5\textwidth, clip]{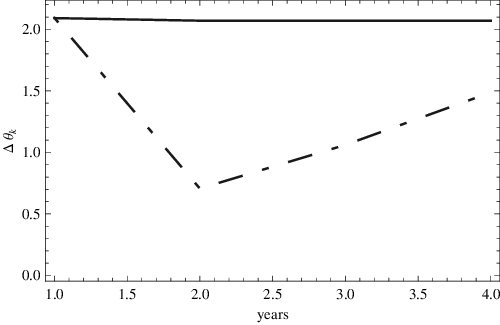}
}
\centerline{
\includegraphics[width=0.5\textwidth, clip]{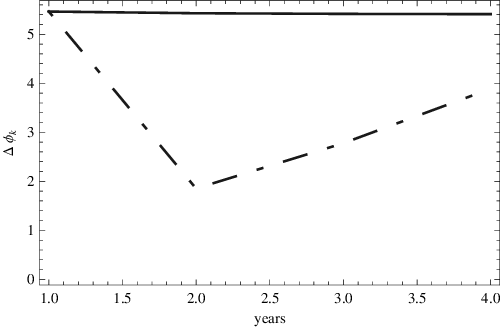}
\includegraphics[width=0.5\textwidth, clip]{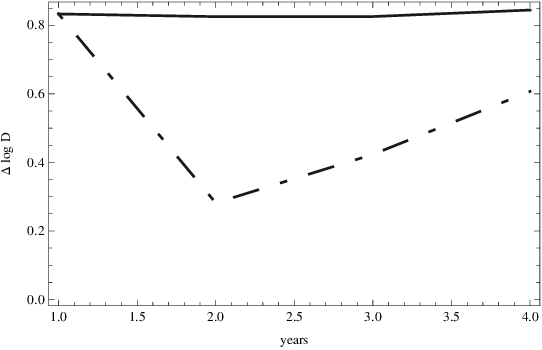}
}
\caption{Same as in Figure \ref{fn1} but for extrinsic parameters.}
\label{fn2}
\end{figure*}

The Tables show estimates for the noise-induced parameter errors if the last year of inspiral is observed. However, LISA may observe a binary system in various stages of evolution. In Figures \ref{fn1} and \ref{fn2} we show how the error estimates vary as a function of the time remaining until plunge at the start of the observation. These results assume a one year LISA observation with fixed parameters $m=10M_{\odot}$, $M=10^6M_{\odot}$, $q=0.9$, \(\Phi_0=0,\theta_S=\pi/4,\phi_S=0,\theta_K=\pi/8,\phi_K=0\). We present the results in two ways, (1) the distance to the source is adjusted to keep the total SNR$=30$ over the observation;  (2)  the distance to the source is fixed and chosen so that the SNR$=30$ over a one year observation starting one year before plunge. 

Our results for black hole systems, $m=10M_{\odot}$, are broadly consistent with existing results in the literature \cite{cutler}. We find typically that an EMRI observation can determine the CO and SMBH masses and the SMBH spin within fractional errors of \(\sim 10^{-4}, 10^{-3.5}\) and $10^{-4.5}$ respectively. We expect to determine the location of the source on the sky to within \(10^{-3}\) steradians, and determine the SMBH spin orientation to within \(10^{-3.5}\) steradians. However, these results are based on a different waveform model. Our model is based on true geodesics of the Kerr space-time, so among other things we have a more accurate frequency at plunge. The various other modifications, such as the inclusion of the conservative self-force corrections hopefully means this model is providing reliable and independent error estimates. One thing that is significantly different in our results is the estimate of LISA's ability to determine parameters for the neutron star and white dwarf systems. Our error estimates for these systems are even smaller than those for the black holes. The improvement arises because, by evolving $L_z$ rather than $p$, the approach to the innermost stable circular orbit (ISCO) shows the proper rapid change in $p$, which carries more information about the intrinsic parameters. This rapid inspiral to ISCO is missed in PN evolutions. In a sense, these results are optimistic as, given the event rate~\cite{lrs}, we are unlikely to observe such systems in the last year before plunge, and the SNR is likely to be lower, while we have normalised all results to SNR$=30$ in this paper. These results are nonetheless interesting.

In the Monte Carlo simulations that used an evolution of \(p\) instead of \(L_z\) (case (b) mentioned above), we found results that were close to the ones quoted here and were also in good agreement with the results obtained by Barack and Cutler's~\cite{cutler}. However, there were subtle differences. In particular, the distribution of the errors, cf., Figure~\ref{mcfmdis}, is much broader when using the 2PN $p$ evolution. In addition, the error estimate for the SMBH spin is an order of magnitude worse using the  2PN $p$ evolution. 

These various results provide reassurance that our estimates from the kludge model and those by Barack \& Cutler \cite{cutler}, may provide a fair reflection of what will be achieved in practice with LISA. Our Monte Carlo results are also the first of their kind for the EMRI problem to appear in the literature and illustrate the spread in errors that arise from randomisation of the extrinsic source parameters. The variation in the parameter determination accuracy as a function of time remaining until plunge indicates that the accuracy of determination of the intrinsic parameters depends significantly on the which part of the inspiral is observed, while the determination of the extrinsic parameters is equally good no matter which part of the inspiral is detected. This makes sense, as most of the orbital evolution takes place during the last year of inspiral and so the signature of the intrinsic parameters is most strongly present there. In contrast, the determination of the extrinsic parameters, such as the position in the sky and orientation of the SMBH spin, comes primarily from the modulations induced in the waveform by the detector motion, and so do not change significantly provided a full year of inspiral is observed.

\section{Model-induced parameter errors}
\label{s5}
The FM gives an estimate of the errors that arise due to noise in the detector. But that is not the only source of error. The kludge we have built is only approximate and hence the kludge waveform that is the best--fit to the data may have different parameters to the true waveform, introducing another parameter error.  We shall call these ``model'' errors. Cutler and Vallisneri \cite{vallisneri} recently introduced a framework that allows us to estimate the magnitude of the model errors. This relies on knowing what the ``true'' signal is, which we do not. However, we can use the formalism to estimate how important for parameter estimation the inclusion of various terms in the waveform model may be. In this section, we look specifically at what effect ignoring the conservative self--force corrections will have on parameter determination.

The noise error scales as \(1/\textrm{SNR}\) (cf. \eqref{29}), but the model errors are independent of SNR. Therefore, it might be the case that for the EMRI systems with highest SNR, theoretical errors could dominate the total parameter--estimation error. In general, if crossing a term out in the model gives rise to a parameter error that is comparable to or smaller than the noise-induced error, then we can safely ignore that term in our search template. We want to estimate theoretically what error would result from omitting conservative corrections from the waveform model, or by using a model with incomplete conservative corrections. So, we take the ``true'' waveform, \(h_{\rm GR}\), to be the kludge waveform including all conservative pieces. We then estimate the model errors when we search for \(h_{\rm GR}\) using templates,  \(h_{\rm AP}\), that include none or only part of the conservative corrections.  

We now briefly review the Cutler and Vallisneri model error formalism~\cite{vallisneri}. We consider two manifolds embedded in the vector spaces of data streams. One of them is covered by the true waveforms \(\{h_{\rm GR}(\theta^i)\}\) and the other one by the approximate waveforms \(\{h_{\rm AP}(\theta^i)\}\). Given a signal \({\bf s} = {\bf h}_\mathrm{GR}(\hat \theta^i) + {\bf n}\), the best fit  \(\theta^i\) is  determined by the condition 
\begin{equation}
\label{36}
\Big( \partial_j \mathbf{h}_\mathrm{AP}(\theta^i) \Big|  
\mathbf{s} - \mathbf{h}_\mathrm{AP}(\theta^i) \Big) = 0.
\end{equation} 
\noindent Furthermore, at first order in the error \(\Delta\theta^i ({\bf n})\equiv \theta^i ({\bf n})- \hat \theta^i\), Eq. \eqref{36} takes the form
\begin{equation}
\Delta\theta^i 
 =  \Big(\Gamma^{-1}(\tbf)\Big)^{ij} \, \Big( \partial_j \hAP(\tbf) \Big| \, \mathbf{n} \Big)
 +  \Big(\Gamma^{-1}(\tbf)\Big)^{ij} \, \Big( \partial_j \hAP(\tbf) \Big| \, \hGR(\ttr) - \hAP(\ttr) \Big) \, ,
\label{37}
\end{equation}
\noindent where the Fisher matrix is evaluated using the approximate waveforms \(\Gamma_{ij}(\tbf) \equiv ( \partial_i \hAP(\tbf)|\partial_j \hAP(\tbf))\). 

Relation \eqref{37} clearly shows us that, at leading order, \(\Delta \theta^i\) is the sum of two contributions. The first one is due to noise in the detector, \(\Delta_n \theta^i\), whereas  the second one, \(\Delta_\mathrm{th}\theta^i\), is the contribution due to the inaccurate waveform. These are given, respectively, by 
\begin{eqnarray}
\Delta_n\theta^i  & =&   \Big(\Gamma^{-1}(\tbf)\Big)^{ij} \Big( \partial_j \hAP(\tbf) \Big| \mathbf{n} \Big),  \qquad
\Delta_\mathrm{th}\theta^i =  \Big(\Gamma^{-1}(\tbf)\Big)^{ij} \Big( \partial_j \hAP(\tbf) \Big| \hGR(\ttr) - \hAP(\ttr)\Big). \label{38.2}
\end{eqnarray}
\noindent If we knew both \( \ttr\) and the noise realization \({\bf n}\) then these equations would allow us to determine \(\tbf\). However, experimentally we are only able to determine the  \(\hAP(\tbf)\) that is the best fit waveform for a given data stream, \(\mathbf{s}\), and we are unsure about the error \(\Delta \theta \equiv \tbf - \ttr\). In addition, we do not know $\hat{\theta}$ in Eq.~\eqref{38.2}. At leading order, we can replace,  \(\hGR(\ttr) - \hAP(\ttr)\) by \(\hGR(\tbf) - \hAP(\tbf)\), obtaining
\begin{equation}
\label{39}
\Delta_\mathrm{th}\theta^i =  \Big(\Gamma^{-1}(\tbf)\Big)^{ij} \Big( \partial_j \hAP(\tbf) \Big| \hGR(\tbf) - \hAP(\tbf)\Big).
\end{equation}
\noindent This relation is both noise and SNR independent. This property, along with the fact that \(\Delta_\mathrm{th}\theta^i\) is not averaged out if the same event is measured by a large number of nearly identical detectors leads us to consider  \(\Delta_\mathrm{th}\theta^i\) as a systematic error.

\newcommand{\hla}{\mathbf{h}}
\newcommand{\tla}{\theta(\lambda)}
\newcommand{\ttl}{\theta_\mathrm{tr}(\lambda)}

Cutler and Vallisneri~\cite{vallisneri} found that the leading order approximation, Eq.~\eqref{39}, was not very good, unless the waveform is re--written in an amplitude-phase form
\begin{equation}
\label{40}
\tilde h^{\alpha}(f) =  A^{\alpha}(f) e^{i \Psi^{\alpha}(f)} \,.
\end{equation}

\noindent The amplitude \(A\) and phase \(\Psi\) are given by
\begin{eqnarray}
A_I &= & \frac{\sqrt{3} \, m}{2 \, D}\sqrt{A^{2}_{+} F_{I,+}^2 + A_{\times}^2 F_{I,\times}^2} \,, \qquad
\Psi_I =2 \phi + \psi_I, \\
A_{II} &= &\frac{\sqrt{3} \, m}{2 \, D} \sqrt{A^{2}_{+} F_{II,+}^2 + A_{\times}^2 F_{II,\times}^2} \,, \qquad
\Psi_{I I}=2 \phi + \psi_{II},
\end{eqnarray}
\noindent where \(A_{+, \times}\) are given by \eqref{15} with \(A_{+}= A_{\times}= 2 p^2 \dot{\phi}^2\), \(F_{\{I, II\,;\,+, \times\}}\)  are given by \eqref{16} and \eqref{17}, and the other various quantities are given by
\begin{eqnarray}
\psi_I &=& \arctan\left(- \frac{A_{\times}F_{I,\times}}{A_{+}F_{I,+}}\right),\qquad
\psi_{II} = \arctan\left(- \frac{A_{\times}F_{II,\times}}{A_{+}F_{II,+}}\right).
\label{phases}
\end{eqnarray}
\noindent The first order approximation to this expression
\begin{equation}
\label{42}
\Delta_\mathrm{th}\theta^i \approx
\big(\Gamma^{-1}(\tbf) \big)^{ij}  \Big( \underbrace{\big[ \Delta {\bf A} + i {\bf A} \Delta \boldsymbol{\Psi} \big] e^{i \boldsymbol{\Psi}}}_{\text{at}\,\theta} \Big| \,
  \partial_j \hAP(\tbf) \Big),
\end{equation}
\noindent was found to give reliable results when compared to more accurately computed error estimates~\cite{vallisneri}, so we use this form again here. Equation~\eqref{42} behaves better than Eq.~\eqref{39} since the difference between two waveforms, \(\hAP(\tbf)  - \hAP(\ttr)\), is not very well approximated by the first term in its Taylor expansion. The differences in both the amplitude and phase of the waveform are individually well approximated by the linear terms in the Taylor series \cite{vallisneri}. In fact, \eqref{39} is reliable only as long as the phase difference between the two waveforms is much less than one radian, i.e. \( \Delta \theta^j \partial_j \boldsymbol{\Psi}_\mathrm{AP}(\tbf) \ll 1\), whereas for \eqref{42} we just require \(\Delta \theta^i \Delta \theta^j \partial_i \partial_j \boldsymbol{\Psi}_\mathrm{AP}(\tbf) \ll 1\). This condition is much less restrictive than the former one.

We can now use equation \eqref{42} to estimate the magnitude of the parameter errors that arise from inaccuracies in the template waveform. At present, accurate waveforms including all first order conservative self--force corrections are not known. We want to estimate how relevant these corrections are for parameter estimation. We can use the kludge model for this, since we have now included conservative corrections to 2PN order and can therefore turn these corrections on and off. This will allow us to compute the ratio \(\cal{R}\) of the model error that arises from omitting the conservative part of the self-force to the Fisher Matrix error for each of the 10 parameters in the model. This ratio will indicate the importance of the conservative corrections for parameter determination. If \(\cal{R}\) \(\lesssim 1\) then the estimates obtained from a model that ignores the conservative piece should still be reliable, but if \(\cal{R}\) \(>>1\) then it is clear that the parameter estimates would not be trustworthy. The ratio of parameter errors to FM errors obtained from a Monte Carlo simulation are summarized in Tables~\ref{tabFMRatBH}, \ref{tabFMRatNS} and \ref{tabFMRatWD}. As before we have evaluated these model-induced parameter errors using expression \eqref{6} to evolve $p$, and using the full Teukolsky fit expression for $\dot{L_z}$, Eq.~\eqref{new_Ldot}. We have also done Monte Carlo simulations with $\dot{L_z}$ truncated at 2PN order, and evolving $p$ at 2PN order directly using Eq.~\eqref{7}. The results were largely consistent between all three simulations.

We quote results for the same three test systems that we used previously, and consider five different comparisons --- for the first two we took the ``true'' waveform to be our kludge waveform with 2PN conservative corrections, and took the template to be a kludge waveform with either no conservative corrections (``$0$PN'') or with conservative corrections to $1.5$PN order (``$1.5$PN''); for the latter three comparisons we used the self-force model in the three varieties listed in Table~\ref{t3} --- (1) $\delta \dot{L_z}=0$ and all $\epsilon_i=0$ in Eq.~\eqref{drdLSF} (``1st order''); (2) $\delta \dot{L_z}=0$ and the $\epsilon$ coefficients set to the values in Eqs.~\eqref{sfco}--\eqref{sfco1} (``incomplete''); (3) as (2) but now with $\delta \dot{L_z}$ from Eq.~\eqref{sfKco1} (``2nd order'') --- and did pairwise comparisons.

\begin{table} [htp]
\begin{tabular}{|c|c|c|c|c|c|c|c|c|c|c|c|}
\hline\multicolumn{2}{|c|}{}&\multicolumn{10}{c|}{Statistics of distribution of \(\log_{10}({\cal{R}})\) for error ratio, \({\cal{R}}\), in parameter \(X =\)}\\\cline{3-12}
\multicolumn{2}{|c|}{Model}&$\ln m$&$\ln M$&$q$&$p_0$&$\phi_0$&$\theta_S$&$\phi_S$&$\theta_K$&$\phi_K$ &$\ln D$\\\hline
&Mean&-0.22&-0.23&-0.17&-0.22&-0.09&-0.20&-0.20&0.00&0.03&0.60\\\cline{2-12}
&St. Dev.&0.584&0.626&0.572&0.611&0.736&0.581&0.534&0.750&0.745&0.832\\\cline{2-12}
Kludge&L. Qt.&-0.59&-0.53&-0.48&-0.54&-0.41&-0.43&-0.47&-0.38&-0.29&0.19\\\cline{2-12}
2PN vs 0PN&Med.&-0.11&-0.10&-0.06&-0.10&0.01&-0.10&-0.09&0.11&0.16&0.75\\\cline{2-12}
&U. Qt.&0.25&0.23&0.25&0.23&0.39&0.20&0.17&0.55&0.54&1.20\\\hline
&Mean&-0.45&-0.43&-0.38&-0.43&-0.33&-0.40&-0.44&-0.23&-0.20&0.36\\\cline{2-12}
&St. Dev.&0.598&0.591&0.540&0.594&0.744&0.578&0.606&0.720&0.733&0.841\\\cline{2-12}
Kludge&L. Qt.&-0.75&-0.72&-0.65&-0.71&-0.58&-0.63&-0.66&-0.53&-0.53&0.03\\\cline{2-12}
2PN vs 1.5PN&Med.&-0.32&-0.31&-0.29&-0.32&-0.20&-0.28&-0.30&-0.09&-0.08&0.54\\\cline{2-12}
&U. Qt.&-0.01&-0.00&0.01&-0.00&0.12&-0.01&-0.01&0.24&0.29&0.91\\\hline
&Mean                &-0.04&0.00 &N/A &-0.13 &-0.41 &-0.60&-0.37 &-0.38 &-0.31 &-0.67\\\cline{2-12}
Self Force&St. Dev.  &0.455&0.454&N/A &0.454 &0.454 &0.594&0.523 &0.553 &0.610 &0.572\\\cline{2-12}
``incomplete''&L. Qt.&-0.20&-0.16&N/A &-0.15 &-0.65 &-0.92&-0.74 &-0.59 &-0.49 &-0.63\\\cline{2-12}
vs&Med.              &0.09&0.06  &N/A &0.05  &-0.35 &-0.48&-0.16 &-0.29 &-0.27 &-0.54\\\cline{2-12}
``1st order''&U. Qt. &0.23&0.28  &N/A &0.27  &-0.25 &-0.25&0.00  &-0.13 &-0.11 &-0.47\\\hline

&Mean                &-0.15 &-0.14 &N/A &-0.16 &-0.23 &-0.26 &-0.26 &-0.05 &-0.08 &-0.05\\\cline{2-12}
Self Force&St. Dev.  &0.446 &0.489 &N/A &0.485 &0.735 &0.375 &0.521 &0.678 &0.639 &0.681\\\cline{2-12}
``2nd order''&L. Qt. &-0.37 &-0.44 &N/A &-0.46 &-0.63 &-0.49 &-0.47 &-0.52 &-0.60 &-0.51\\\cline{2-12}
vs&Med.              &-0.15 &-0.11 &N/A &-0.13 &-0.11 &-0.25 &-0.07 &0.02  &0.00  &-0.02\\\cline{2-12}
``1st order''&U. Qt. &0.16  &0.21  &N/A &0.34  &0.34  &0.02  &0.11  &0.45  &0.47  &0.50\\\hline

&Mean                &0.11 &0.08 &N/A  &0.04 &0.13  &0.14  &0.06  &0.28  &0.33  &0.22\\\cline{2-12}
Self Force&St. Dev.  &0.553&0.619&N/A  &0.625&0.747 &0.531 &0.639 &0.705 &0.717 &0.863\\\cline{2-12}
``2nd order''&L. Qt. &-0.26&-0.30&N/A  &-0.36&-0.26 &-0.20 &-0.37 &-0.20 &-0.17 &-0.30\\\cline{2-12}
vs&Med.              &0.17 &0.08 &N/A  &0.09 &0.20  &0.16  &0.21  &0.32  &0.43  &0.35\\\cline{2-12}
``incomplete''&U. Qt.&0.46 &0.47 &N/A  &0.46 &0.66  &0.53  &0.50  &0.85  &0.87  &0.84\\\hline
\end{tabular}
\caption{Summary of Monte Carlo simulation results for the ratio of model errors to Fisher Matrix errors for black hole systems ($m=10M_{\odot}$) using the Teukolsky fit expression for d$L_z$/dt. We show the mean, standard deviation, median and quartiles of the distribution of the logarithm to base ten of the ratio for each parameter. Results are given for various comparisons, as indicated and described in the text. A comparison ``A vs B'' uses model A as the true waveform, and model B as the search template.}
\label{tabFMRatBH}
\end{table}

\begin{table}[ht]\small
\begin{tabular}{|c|c|c|c|c|c|c|c|c|c|c|c|}
\hline\multicolumn{2}{|c|}{}&\multicolumn{10}{c|}{Statistics of distribution of \(\log_{10}({\cal{R}})\) for error ratio, \({\cal{R}}\), in parameter \(X =\)}\\\cline{3-12}
\multicolumn{2}{|c|}{Model}&$\ln m$&$\ln M$&$q$&$p_0$&$\phi_0$&$\theta_S$&$\phi_S$&$\theta_K$&$\phi_K$ &$\ln D$\\\hline
&Mean&-0.43&-0.35&-0.33&-0.36&-0.35&-0.33&-0.36&-0.23&-0.21&0.37\\\cline{2-12}
&St. Dev.&0.563&0.587&0.586&0.626&0.728&0.589&0.601&0.728&0.754&0.832\\\cline{2-12}
Kludge&L. Qt.&-0.75&-0.62&-0.60&-0.62&-0.59&-0.55&-0.58&-0.48&-0.44&0.05\\\cline{2-12}
2PN vs 0PN&Med.&-0.30&-0.20&-0.19&-0.20&-0.19&-0.18&-0.20&-0.10&-0.04&0.57\\\cline{2-12}
&U. Qt.&-0.00&0.06&0.08&0.06&0.08&0.07&0.03&0.23&0.23&0.89\\\hline
&Mean&-0.80&-0.67&-0.68&-0.68&-0.73&-0.68&-0.71&-0.57&-0.56&0.02\\\cline{2-12}
&St. Dev.&0.576&0.551&0.657&0.573&0.744&0.598&0.623&0.711&0.760&0.817\\\cline{2-12}
Kludge&L. Qt.&-1.09&-0.90&-0.88&-0.90&-0.98&-0.90&-0.91&-0.83&-0.77&-0.23\\\cline{2-12}
2PN vs 1.5PN&Med.&-0.66&-0.53&-0.50&-0.54&-0.57&-0.51&-0.56&-0.43&-0.40&0.21\\\cline{2-12}
&U. Qt.&-0.37&-0.29&-0.28&-0.29&-0.31&-0.26&-0.33&-0.15&-0.12&0.52\\\hline
&Mean                &-0.90&-0.84&N/A&-0.84&-1.02&-1.11&-1.08&-1.23&-1.19&-1.21\\\cline{2-12}
Self Force&St. Dev.  &0.456&0.403&N/A&0.404&0.642&0.543&0.521&0.563&0.569&0.544\\\cline{2-12}
``incomplete''&L. Qt.&-1.06&-1.01&N/A&-1.03&-1.25&-1.37&-1.23&-1.41&-1.44&-1.32\\\cline{2-12}
vs&Med.              &-0.85&-0.78&N/A&-0.82&-0.92&-1.08&-0.99&-1.17&-1.15&-1.16\\\cline{2-12}
``1st order''&U. Qt. &-0.62&-0.58&N/A&-0.57&-0.69&-0.93&-0.81&-0.94&-0.95&-0.91\\\hline

&Mean                &-0.95&-0.98&N/A&-0.99&-1.13&-1.37&-1.23&-1.90&-1.88&-1.76\\\cline{2-12}
Self Force&St. Dev.  &0.556&0.557&N/A&0.584&0.680&0.571&0.507&0.672&0.614&0.580\\\cline{2-12}
``2nd order''&L. Qt. &-1.14&-1.07&N/A&-1.09&-1.47&-1.77&-1.48&-2.19&-2.27&-1.97\\\cline{2-12}
vs&Med.              &-0.89&-0.94&N/A&-0.91&-1.04&-1.23&-1.19&-1.76&-1.75&-1.69\\\cline{2-12}
``1st order''&U. Qt. &-0.73&-0.69&N/A&-0.79&-0.85&-0.89&-0.91&-1.41&-1.44&-1.42\\\hline

&Mean                &-0.76&-0.82&N/A&-0.84&-0.83&-0.84&-0.80&-0.66&-0.63&-0.69\\\cline{2-12}
Self Force&St. Dev.  &0.434&0.603&N/A&0.626&0.697&0.572&0.634&0.716&0.732&0.803\\\cline{2-12}
``2nd order''&L. Qt. &-0.98&-1.20&N/A&-1.20&-1.19&-1.09&-1.10&-0.99&-0.92&-0.99\\\cline{2-12}
vs&Med.              &-0.64&-0.70&N/A&-0.75&-0.73&-0.69&-0.66&-0.53&-0.47&-0.56\\\cline{2-12}
``incomplete''&U. Qt.&-0.39&-0.29&N/A&-0.38&-0.36&-0.48&-0.41&-0.16&-0.12&-0.21\\\hline
\end{tabular}
\caption{As Table~\ref{tabFMRatBH}, but for neutron star inspirals ($m=1.4M_{\odot}$).}
\label{tabFMRatNS}
\end{table}

\begin{table} [ht]
\begin{tabular}{|c|c|c|c|c|c|c|c|c|c|c|c|}
\hline\multicolumn{2}{|c|}{}&\multicolumn{10}{c|}{Statistics of distribution of \(\log_{10}({\cal{R}})\) for error ratio, \({\cal{R}}\), in parameter \(X =\)}\\\cline{3-12}
\multicolumn{2}{|c|}{Model}&$\ln m$&$\ln M$&$q$&$p_0$&$\phi_0$&$\theta_S$&$\phi_S$&$\theta_K$&$\phi_K$ &$\ln D$\\\hline
&Mean&-0.37&-0.52&-0.52&-0.53&-0.55&-0.49&-0.52&-0.43&-0.39&0.17\\\cline{2-12}
&St. Dev.&0.526&0.589&0.661&0.614&0.782&0.657&0.643&0.786&0.772&0.878\\\cline{2-12}
Kludge&L. Qt.&-0.57&-0.72&-0.70&-0.73&-0.78&-0.69&-0.71&-0.65&-0.60&-0.07\\\cline{2-12}
2PN vs 0PN&Med.&-0.22&-0.38&-0.36&-0.38&-0.38&-0.31&-0.35&-0.25&-0.22&0.41\\\cline{2-12}
&U. Qt.&0.00&-0.13&-0.12&-0.13&-0.11&-0.07&-0.12&0.03&0.06&0.71\\\hline
&Mean&-0.71&-0.85&-0.86&-0.86&-0.96&-0.87&-0.86&-0.81&-0.79&-0.22\\\cline{2-12}
&St. Dev.&0.491&0.518&0.589&0.530&0.781&0.632&0.628&0.790&0.769&0.870\\\cline{2-12}
Kludge&L. Qt.&-0.88&-1.06&-1.05&-1.06&-1.20&-1.09&-1.00&-1.04&-1.00&-0.48\\\cline{2-12}
2PN vs 1.5PN&Med.&-0.59&-0.76&-0.74&-0.76&-0.79&-0.70&-0.71&-0.65&-0.63&-0.01\\\cline{2-12}
&U. Qt.&-0.38&-0.49&-0.48&-0.49&-0.51&-0.46&-0.48&-0.33&-0.34&0.32\\\hline
&Mean                &-1.27&-1.21&N/A&-1.22&-1.32&-1.34&-1.30&-1.19&-1.15&-1.22\\\cline{2-12}
Self Force&St. Dev.  &0.596&0.586&N/A&0.591&0.797&0.608&0.601&0.792&0.720&0.773\\\cline{2-12}
``incomplete''&L. Qt.&-1.54&-1.57&N/A&-1.54&-1.62&-1.55&-1.52&-1.43&-1.42&-1.56\\\cline{2-12}
vs&Med.              &-1.16&-1.13&N/A&-1.14&-1.19&-1.24&-1.16&-1.02&-1.05&-1.05\\\cline{2-12}
``1st order''&U. Qt. &-0.83&-0.73&N/A&-0.74&-0.91&-0.93&-0.85&-0.70&-0.68&-0.61\\\hline

&Mean                &-1.65&-1.55&N/A&-1.65&-2.19&-2.22&-2.12&-2.05&-2.02&-1.96\\\cline{2-12}
Self Force&St. Dev.  &0.425&0.442&N/A&0.447&0.581&0.523&0.537&0.645&0.575&0.683\\\cline{2-12}
``2nd order''&L. Qt. &-1.79&-1.67&N/A&-1.79&-2.60&-2.57&-2.49&-2.47&-2.45&-2.25\\\cline{2-12}
vs&Med.              &-1.59&-1.47&N/A&-1.55&-2.11&-2.07&-1.99&-1.94&-1.90&-1.99\\\cline{2-12}
``1st order''&U. Qt. &-1.36&-1.30&N/A&-1.37&-1.61&-1.75&-1.66&-1.49&-1.58&-1.47\\\hline

&Mean                &-1.24&-1.31&N/A&-1.32&-1.34&-1.37&-1.33&-1.19&-1.14&-1.22\\\cline{2-12}
Self Force&St. Dev.  &0.454&0.519&N/A&0.604&0.697&0.541&0.589&0.721&0.651&0.753\\\cline{2-12}
``2nd order''&L. Qt. &-1.45&-1.61&N/A&-1.61&-1.64&-1.58&-1.60&-1.47&-1.42&-1.54\\\cline{2-12}
vs&Med.              &-1.19&-1.23&N/A&-1.26&-1.21&-1.26&-1.20&-1.05&-1.04&-1.06\\\cline{2-12}
``incomplete''&U. Qt.&-0.94&-0.85&N/A&-0.86&-0.94&-1.02&-0.90&-0.76&-0.74&-0.75\\\hline
\end{tabular}
\caption{As Table~\ref{tabFMRatBH}, but for white dwarf inspirals ($m=0.6M_{\odot}$).}
\label{tabFMRatWD}
\end{table}

\begin{figure*}[ht]
\centerline{
\includegraphics[height=0.5\textwidth,angle=270,  clip]{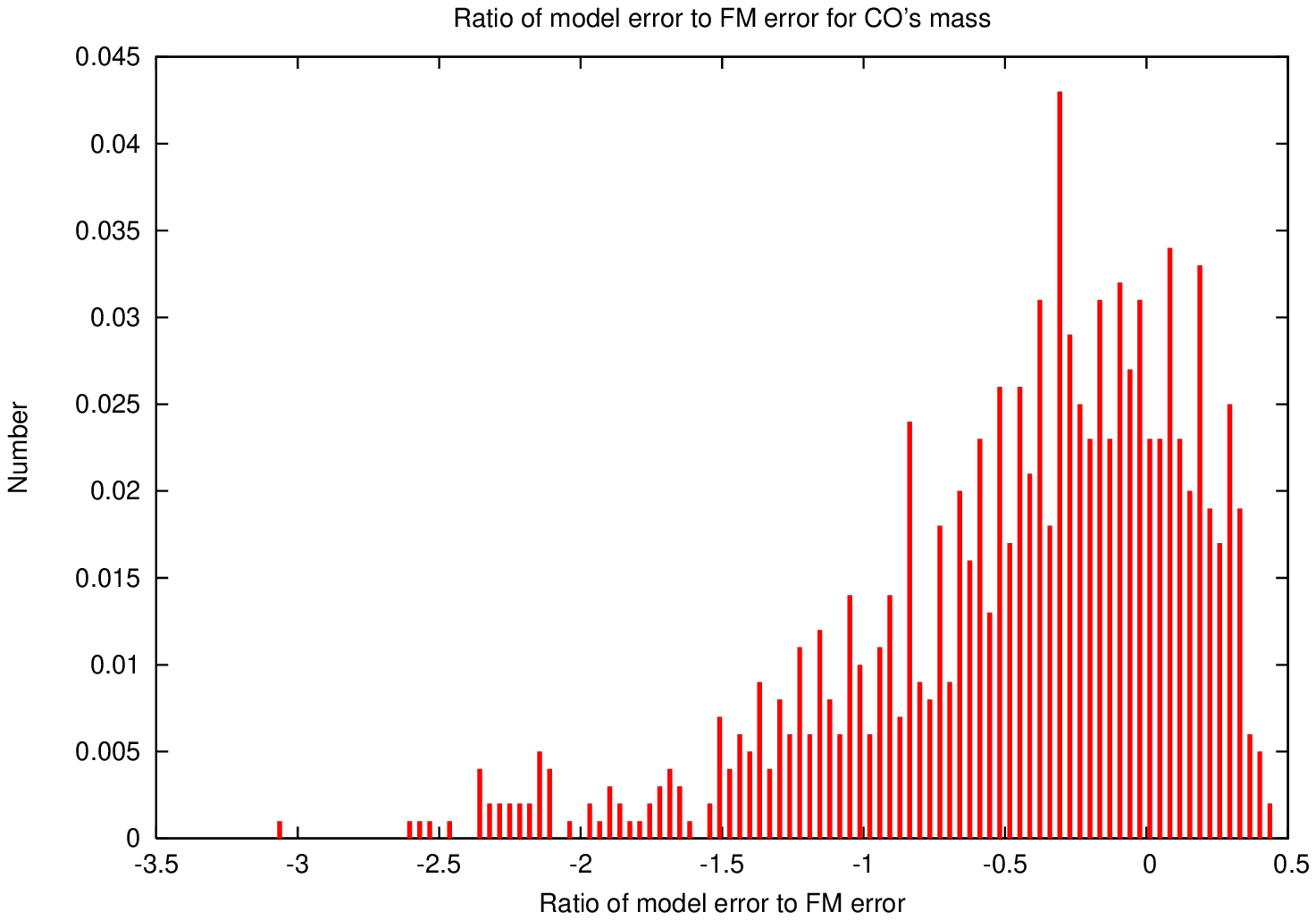}
\includegraphics[height=0.5\textwidth,angle=270, clip]{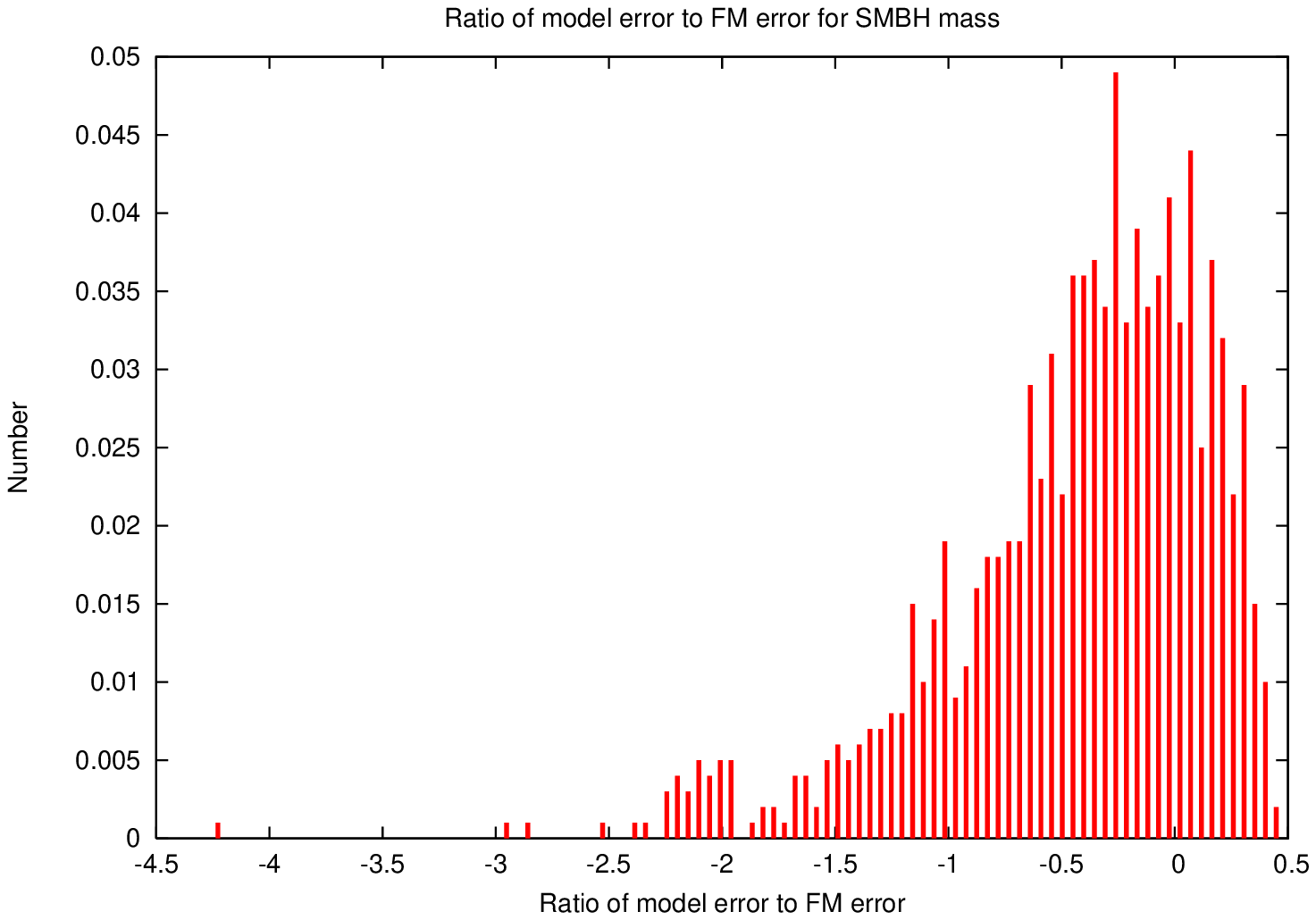}
}
\centerline{
\includegraphics[height=0.5\textwidth,angle=270, clip]{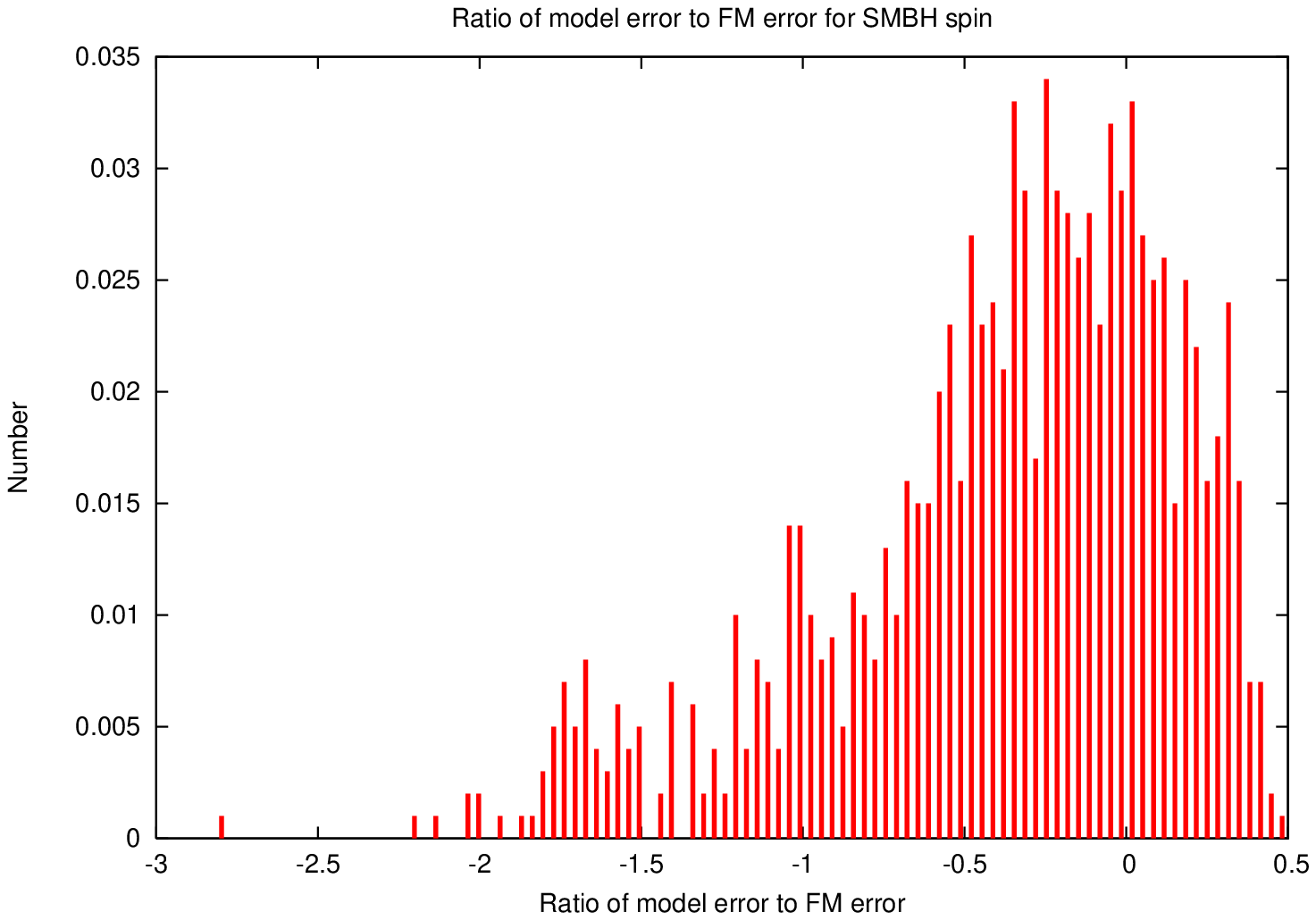}
\includegraphics[height=0.5\textwidth,angle=270, clip]{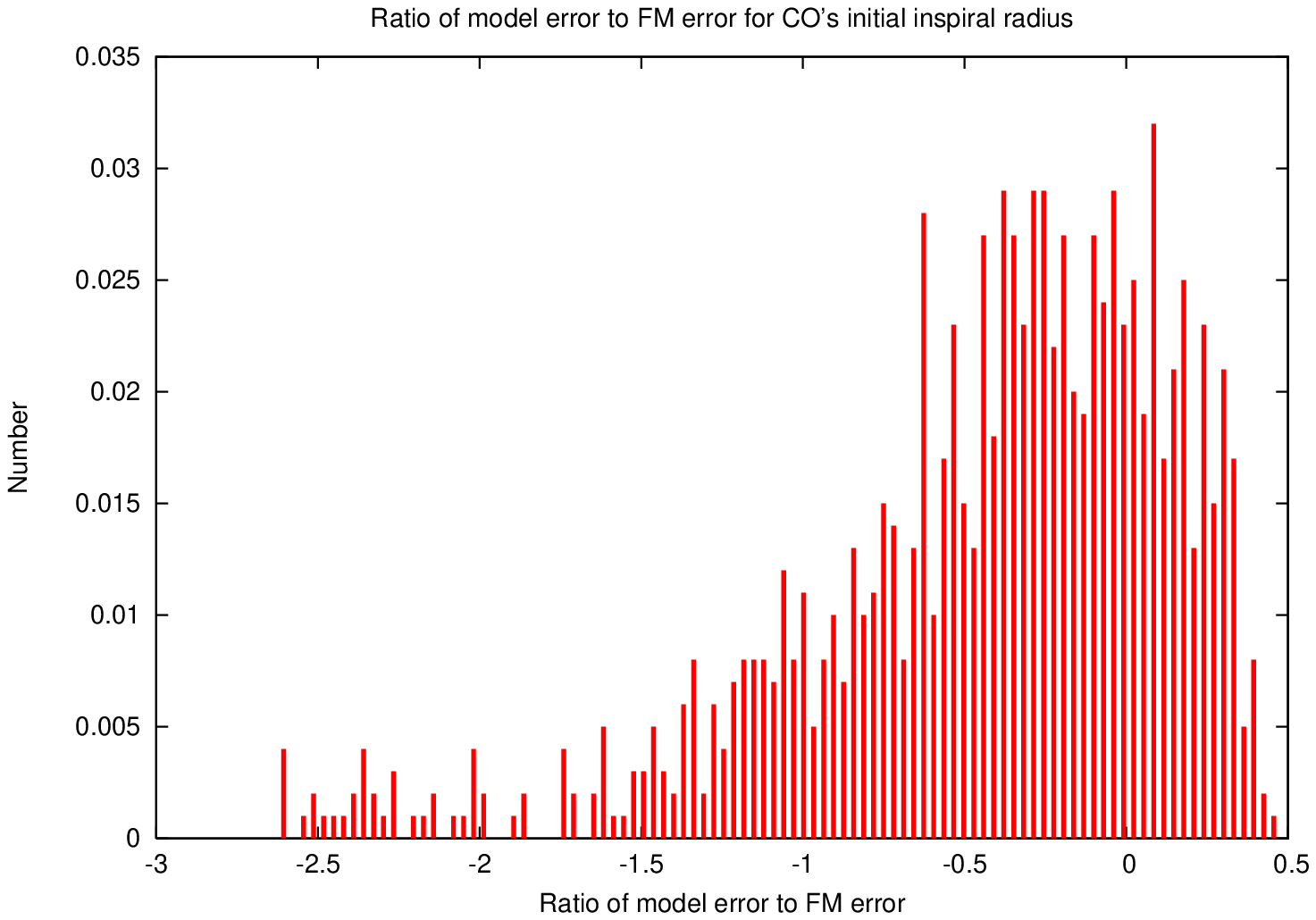}
}
\caption{ Distribution of the ratio of the model induced error to the noise induced FM error for the intrinsic parameters. This data is for the black hole inspiral system ($m=10M_{\odot}$) for the kludge 2PN vs 1.5PN comparison.}
\label{raterrdis}
\end{figure*}

In Figure~\ref{raterrdis}, we also show the full distribution of the error ratio, \(\cal{R}\), for the intrinsic parameters, as computed from the Monte Carlo simulation. These are shown for the black hole systems in the kludge 2PN vs 1.5PN comparison. These plots show that, in this particular case, the vast majority of sources fulfill the condition that the model errors are smaller than the noise induced errors. For those binary systems that do not fulfill this condition, the upper bound on the error ratio \(\cal{R}\)  is \(\cal{R}\) \(<3\). Tables~\ref{tabFMRatBH}, \ref{tabFMRatNS} and \ref{tabFMRatWD} indicate that, for both kludge model comparisons, the ratio \(\cal{R}\) \(<1\) for all parameters at most points in the parameter space for all three types of inspiral. In fact, for any of the model parameters, less than \(0.15\%\) of points in the Monte Carlo runs had \(2<\cal{R}\) \(<3\). These results suggest that including conservative corrections is not essential for accurate parameter determination, but including them up to 2PN order will certainly reduce the model errors further.

Things are not so clear cut for the self-force comparisons, however. For the neutron star and white dwarf inspirals, the ratio \(\cal{R}\) \(<1\) for practically every element of the parameter space. In fact, for any parameter, less than \(0.1\%\) of points of the Monte Carlo runs satisfy \(2<\cal{R}\) \(<3\). For black holes, the model errors appear generally larger, with ratios typically greater than $1$ in the comparisons to the ``2nd order'' model that was built from the kludge. One thing that must be born in mind for these comparisons is that the kludge model was constructed by comparison to a weak-field post-Newtonian expansion, but we are now comparing it to fully accurate self-force computations in the strong-field. Thus, the reason for the apparently larger discrepancy may be that the kludge model is itself not accurate enough in the regime where we are comparing. This was the reason that we also did the ``incomplete'' vs ``1st order'' comparisons for the self-force models. These expressions do not use the kludge model, but are in the spirit of the other kludge comparisons in that we are crossing out the last term in the model and looking at what effect this has. Those results show generally smaller errors, with more than $50\%$ of points having \(\cal{R}\) \(< 1\). The conclusion is that, in the worst case, a template that omits conservative corrections will identify parameters that are (conservatively) within $10$ Fisher Matrix errors of the true parameters, and this region of parameter space can then be followed up using more accurate waveforms if these are available. We also conclude that the kludge model with 2PN conservative corrections is a reasonably good approximation to the self-force model, since although the ratios for those comparisons are larger, they are still manageably small.

The last two rows of the tables also indicate that it may be better to work consistently to a certain order in mass ratio, rather than to include an incomplete term at higher order. The ``incomplete'' model includes second order in $\eta$ corrections to the evolution of the orbital radius that arise from the first order conservative part of the self-force, but not those that arise from the second order radiative part. The ``1st order'' model includes no second order corrections to the rate of change of radius. The tables indicate that the ``1st order'' model actually leads to smaller parameter errors (when searching for the kludge waveform) than the ``incomplete'' model, although the difference between the two cases is relatively small.

One other thing that we should point out is that we are normalising to SNR$=30$ for all the results. This is an estimate of the threshold SNR that will be required for EMRI detection~\cite{seoane}, but nearby events may have SNR as high as several hundred~\cite{emrirate}. For such systems, the noise induced errors will be smaller than those quoted here by a factor of SNR$/30$, while the model induced errors will be the same. Thus, for accurate parameter determination for the loudest sources LISA detects, these results suggest that it will almost certainly be necessary to include first-order conservative and, if possible, second-order radiative terms in the model.

\begin{figure*}[htp]
\centerline{
\includegraphics[width=0.5\textwidth,  clip]{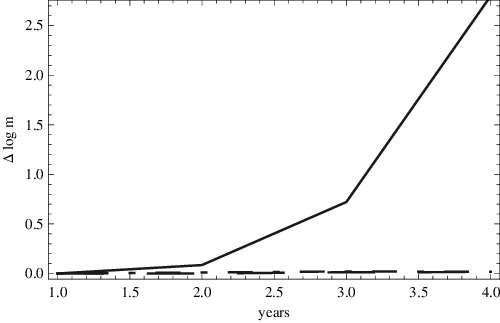}
\includegraphics[width=0.5\textwidth, clip]{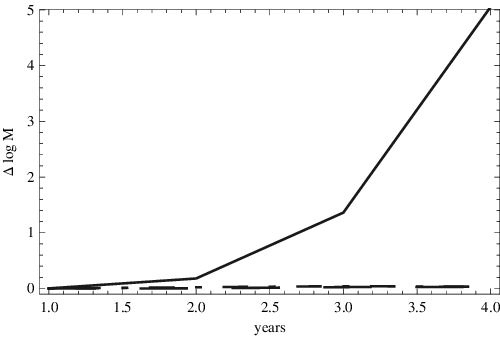}
}
\centerline{
\includegraphics[width=0.5\textwidth,  clip]{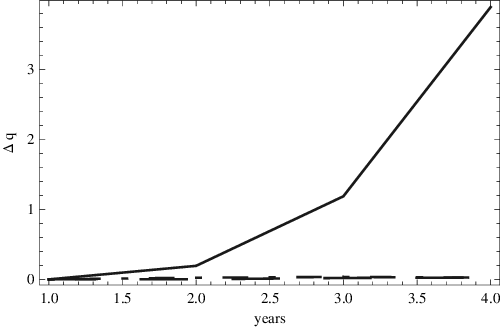}
\includegraphics[width=0.5\textwidth, clip]{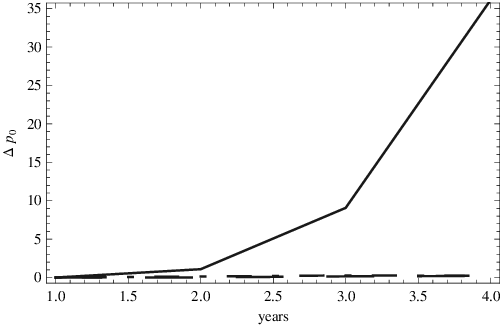}
}
\caption{ We show how the noise induced and model errors vary in a one year observation as a function of the time remaining until plunge at the start of the observation. We show noise induced errors (solid lines), and model errors for two cases, namely (\(h_{\textrm{GR}},h_{\textrm{AP}}\))= (2PN, 0PN) (dashed lines), and (\(h_{\textrm{GR}},h_{\textrm{AP}}\))= (2PN, 1.5PN) (dot-dash lines). This is for a  \( 10M_{\odot}\) CO inspiralling into a  \( 10^6 M_{\odot}\) SMBH with parameter spin \(q=0.9\). We have set the various extrinsic parameters as follows: \(\Phi_0 =0\), \(\theta_S= \pi/4\), \(\phi_S=0\), \(\theta_K=\pi/8\), \(\phi_K=0\). The noise induced errors have been re-normalised to a signal to noise ratio of $30$ over the year of observation. These four plots show the errors in the intrinsic parameters of the system.}
\label{figyears1}
\end{figure*}

\begin{figure*}[htp]
\centerline{
\includegraphics[width=0.5\textwidth,  clip]{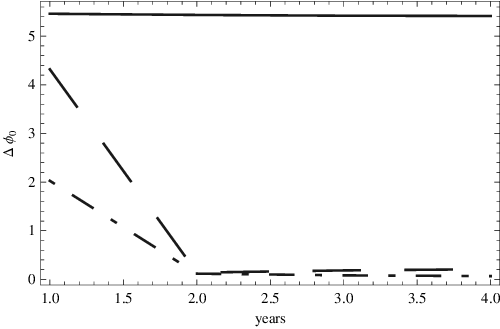}
\includegraphics[width=0.5\textwidth, clip]{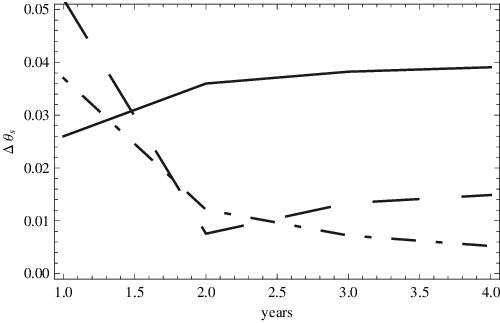}
}
\centerline{
\includegraphics[width=0.5\textwidth,  clip]{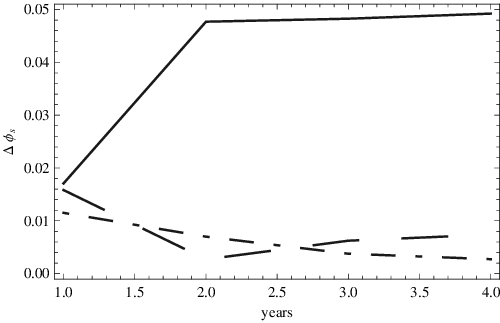}
\includegraphics[width=0.5\textwidth, clip]{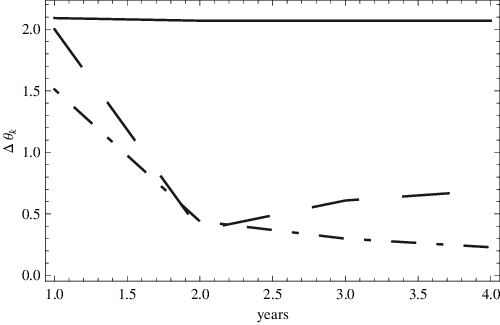}
}
\centerline{
\includegraphics[width=0.5\textwidth, clip]{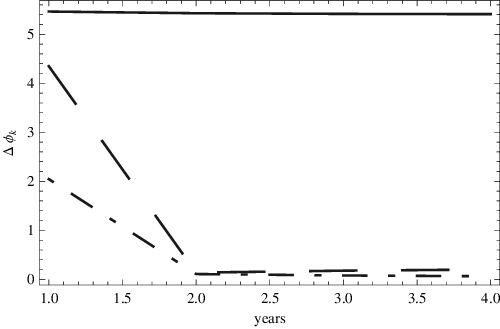}
\includegraphics[width=0.5\textwidth, clip]{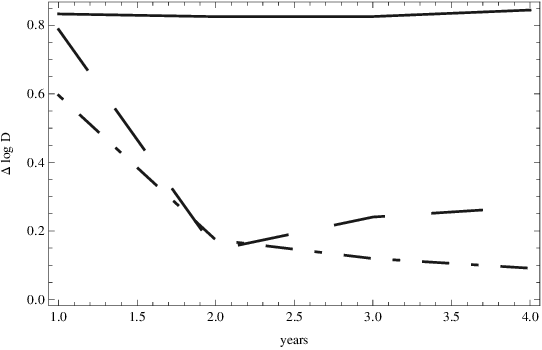}
}
\caption{As Figure \ref{figyears1} but now we show the errors in the extrinsic parameters.}
\label{figyears2}
\end{figure*}

In Figures~\ref{figyears1} and \ref{figyears2} we show how the model and FM errors vary for a one year observation as a function of the time remaining until plunge at the start of the observation. This is for a black hole inspiral, $m=10M_{\odot}$, with the following fixed values of the extrinsic parameters, \(\Phi_0=0,\theta_S=\pi/4,\phi_S=0,\theta_K=\pi/8,\phi_K=0\). In each plot, the noise errors are weighted by a factor \(\textrm{SNR}/30\) to ensure the SNR is $30$ over the one year observation. We have also produced similar plots with the source kept at a fixed distance, but we do not show these here as they do not look very different. 

\clearpage

From Figures \ref{figyears1} and \ref{figyears2} we see that, for this typical source, (1) the noise induced errors in the intrinsic parameters increase quickly if we observe only an earlier part of the inspiral, while the model errors only increase by a small amount. Hence, we would only expect to be model-error dominated when observing the last year if at all; (2) noise and model errors for the extrinsic parameters do not change substantially as we vary the initial time; (3) the theoretical errors decrease as we include more conservative corrections.

It is worth pointing out that the results quoted in Tables~\ref{tabFMRatBH}, \ref{tabFMRatNS}, and \ref{tabFMRatWD} were obtained from Fisher Matrices which were nicely convergent over several orders of magnitude in the offset used to compute the numerical waveform derivatives.  The noise-induced and model errors obtained from the inverses of these matrices also exhibited good convergent behaviour. This is a relevant observation as the matrices we are dealing with have very large condition numbers --- the ratio of the largest to smallest eigenvalues. Typically, we found that using an LU decomposition, i.e., writing the FM as the product of a lower triangular matrix and an upper triangular matrix, the inverse Fisher matrices converged to $\lesssim8\%$ over three orders of magnitude in the numerical offsets.

\section{Conclusions}
\label{s6}
We have constructed an improved kludge model of gravitational wave emission from circular-equatorial EMRIs by including conservative self-force corrections up to 2PN order. We have obtained a 2PN expression for \(\dot \Omega\) which includes  both the conservative self-force at first order in \(\eta\), and quadratic terms of the spin parameter \(q\) at 2PN order. Previous expressions included either conservative corrections \cite{blanchet} or quadratic terms of the spin parameter \cite{tanaka}, but not both. Our model has shown us that the inclusion of conservative corrections has a relatively small impact on the waveform phasing, and so are not essential for source detection, but it will be useful to include them for parameter estimation. At this PN order, the model provides parameter determination accuracy estimates for black hole inspiral systems, $m=10M_{\odot}$, that are broadly consistent with previous results by Barack and Cutler \cite{cutler}. For a typical source at SNR of $30$, we find that a LISA EMRI observation should be able to determine the CO and SMBH masses and the SMBH spin magnitude to within fractional errors of \(\sim 10^{-4}, 10^{-3.5}\) and  \(10^{-4.5}\) respectively. We should also be able to determine the location of the source on the sky to within \(10^{-3}\) steradians, and determine the SMBH spin orientation to within \(10^{-3.5}\) steradians. Our model should be more reliable in the strong field regime as we have built it using a true Kerr geodesic, have included conservative corrections in a physically consistent way and have evolved the geodesic parameters using the best available radiative flux, $\dot{L_z}$. The fact that we get such consistent results provides reassurance that both estimates will be a reasonable reflection of the precision that LISA will ultimately achieve. We have also obtained new results for neutron star and white dwarf inspirals that indicate LISA will also be able to return highly accurate parameters for these systems, provided that the last year of inspiral is observed.

We have also studied in detail the importance of the first order conservative part and the second-order radiative part of the self-force for parameter estimation accuracy using the formalism of Cutler and Vallisneri~\cite{vallisneri}.  We have found that, for these sources, the model errors that arise from omitting these self-force terms are generally smaller than the parameter errors that arise from instrumental noise when the source has SNR$=30$. In our Monte Carlo simulations, no points had model-error to noise-error ratios, $\cal{R}$, greater than $3$ and less than \(0.15\%\) of the points in the Monte Carlo runs lay in the range \(2<\cal{R}\) \(<3\).

We have also compared our results to recently published self-force calculations, that include all first order terms, but nothing at higher order. Comparing to these results allowed us to assess the relative importance of the first order conservative and second order radiative parts of the self-force, as these affect the orbital evolution in the same way. These clearly showed that the missing terms were not necessary for accurate parameter estimation for the inspirals of neutron stars or white dwarfs. However, the results for the inspirals of black holes were less conclusive --- model errors were typically a few times the expected parameter errors from instrumental noise. Part of the reason for this may have been that we derived the kludge corrections by comparison to post-Newtonian expressions in the weak field and then used them to test the self-force results in the strong field. Comparing the self-force waveform to a truncated version of the same did indicate slightly smaller error ratios, although these were still somewhat bigger than in the calculations based entirely on the kludge waveforms.

Our results are the first attempt to assess the necessity of including conservative corrections in templates for parameter estimation with LISA. The results are not absolutely conclusive since they show neither that the model errors are always completely negligible nor that the model errors always overwhelm the errors from instrumental noise. Instead, we find that the two errors are generally comparable. This suggests that search templates can certainly ignore conservative corrections, but it may be necessary to follow-up with more accurate templates to get more precise parameter estimates. The potential problem is that our results suggest the second order radiative part of the self-force may be as important as the first order conservative piece and it is not clear whether templates that include both will be available on the timescale of LISA. Nonetheless, for all the cases we have considered, the model errors are only a few times the noise errors which means that a) we will obtain good estimates of the source parameters, although quoted error bounds must allow for the model error; b) if more accurate templates are available, we will only need to use these (presumably computationally more expensive templates) in an area of parameter space approximately ten times the error box predicted by the Fisher Matrix.

These results are not the full story, since we have considered a rather special class of EMRIs, namely those that are both circular and equatorial. If non-standard EMRI channels operate efficiently, these may make up a significant fraction of LISA events~\cite{seoane}. However, we expect orbits generically to be both eccentric and inclined to the equatorial plane. Although we hope our results are representative of the general case, this needs to be properly explored in the future. It is relatively straightforward to adapt the formalism presented here to the generic case. In general, rather than having just one frequency and frequency derivative, an orbit can be characterised by three frequencies and their derivatives (the perihelion and orbital plane precession frequencies in addition to the orbital frequency) and described by three parameters, e.g., a semi-latus rectum, eccentricity and inclination. Comparison of these six quantities between the kludge and post-Newtonian formalisms will allow determination of the coordinate transformation that relates the two systems and also the computation of the conservative corrections to the three frequencies. This is work that should be done in the near future, to verify the conclusions of this paper in a more general context.

\section*{Acknowledgments}
We thank Steve Drasco and Leor Barack for useful discussions. E.H. is supported by CONACyT. J.G.'s work is supported by the Royal Society.

\bibliography{references}

\end{document}